\documentclass[journal=jacsat,manuscript=article]{achemso}

 \pdfoutput=1

\setkeys{acs}{maxauthors = 0}      % will list all authors

\usepackage{chemformula} % Formula subscripts using \ch{}
\usepackage[T1]{fontenc} % Use modern font encoding
\usepackage{graphicx}
\usepackage{amsmath}
\usepackage{chngcntr}
\usepackage{mathtools}
\usepackage{amssymb}
\usepackage{textgreek}

\setkeys{acs}{articletitle=true}
\makeatletter
\newcommand*{\forcekeywords}{
  \acs@keywords@print
  \let\acs@keywords@print\relax
}
\makeatother

\author{Cyrille Armel Sayou Ngomsi}
\affiliation[Howard University] {Department of Physics and Astronomy, Howard University, Washington, D.C. 20059, USA}
\author{Sai Krishna Narayanan}
\affiliation[University of Maryland]
{Department of Physics, University of Maryland, College Park, Maryland 20742, USA}
\author{Pratibha Dev}
\email{pdev@lps.umd.edu}
\affiliation[Laboratory for Physical Sciences]
{Laboratory for Physical Sciences, College Park, Maryland 20740, United States}
\alsoaffiliation[Howard University]
{Department of Physics and Astronomy, Howard University, Washington, D.C. 20059, USA}

\title[\texttt{achemso} demonstration]{R\MakeLowercase{obust} A\MakeLowercase{luminum} N\MakeLowercase{itride} P\MakeLowercase{assivation} of S\MakeLowercase{ilicon} C\MakeLowercase{arbide} with N\MakeLowercase{ear}-S\MakeLowercase{urface} Q\MakeLowercase{uantum} E\MakeLowercase{mitters} for Q\MakeLowercase{uantum} C\MakeLowercase{omputing} and S\MakeLowercase{ensing} A\MakeLowercase{pplications}}


\begin{document}

\begin{abstract}

Silicon carbide (SiC) hosts a number of point defects that are being explored as single-photon emitters for quantum applications.  Unfortunately, these quantum emitters lose their photostability when placed in proximity to the surface of the host semiconductor. 
In principle, a uniform passivation of the surface's dangling bonds by simple adsorbates, such as hydrogen or mixed hydrogen/hydroxyl groups, should remove detrimental surface effects. However, the usefulness of atomic and molecular passivation schemes is limited by their lack of long-term chemical and/or thermal stability. In this first principles work, we use aluminum nitride (AlN) to passivate SiC surfaces in a core-shell nanowire model. By using a negatively charged silicon vacancy in SiC as the proof-of-principle quantum emitter, we show that AlN-passivation is effective in removing SiC surface states from the band gap and in restoring the defect's optical properties. We also report the existence of a silicon vacancy-based defect at the SiC-AlN interface, which displays distinct spin and optical properties as compared to the other well-studied defects in SiC.

\end{abstract}
\forcekeywords

\section*{Introduction}
Quantum bits (qubits) and single photon emitters are essential building blocks for quantum information science applications.  Wide band gap semiconductors with fluorescent, spin-active point defects can furnish both qubits and single photon emitters for quantum technologies. This is particularly true for different bright defects in silicon carbide (SiC), which not only emit single photons in the favorable near-telecom range~\cite{Christle2015,Fuchs2015}, but they also possess addressable spins~\cite{falk2013polytype,widmann2015coherentRT,Seo2016,nagy2019highfidelity} with long coherence times~\cite{Seo2016,Christle2015,Carter2015,nagy2019highfidelity,widmann2015coherentRT}. The host material itself offers additional advantages, such as low cost, commercial availability, and ease of nanofabrication, all of which can enable scalable SiC-based quantum technologies. 

In order to enhance the radiative signal from defects for quantum computing and/or enhance their sensitivity for quantum sensing applications, different strategies are used, such as shallow implantation and/or nanofabrication of optical cavities in the host materials~\cite{gadalla2021enhanced,majety2021quantum,crookEHu2020purcell, bracherEHu2017selective,nagy2018quantumdichroic,lukinRadulaski20204h,radulaski2017scalable}. A consequence of these strategies is the near-surface placement of the defect, which inadvertently modulates its properties. This has been reported in experiments~\cite{lukinRadulaski20204h,YuanDeLeon2020Charge,sangtawesinDeLeon2019origins,nanostr_VSi_expt_2020,neethirajan2023} and explored in a few theoretical works~\cite{Kaviani2014,LiGali2019,Lofgren_2019,Joshi2022,Sayou_PRMater2024}. 
In particular, two recent theoretical works~\cite{Joshi2022,Sayou_PRMater2024} studied properties of the near-surface, negatively-charged silicon monovacancy ($\mathrm{V_{Si}^{-1}}$) in a 2H-SiC nanowire (NW).  In their first principles study of defects in an unpassivated NW, Joshi and Dev~\cite{Joshi2022} demonstrated that the changes to the frequency of quantum emission from $\mathrm{V_{Si}^{-1}}$ and the loss of the defect's photostability result from the spatial and energetic proximity of the surface states to the defect states.  Sayou Ngomsi \textit{et al.}~\cite{Sayou_PRMater2024} further showed that the aforementioned detrimental surface effects can be remedied by passivation of the SiC surfaces with simple adsorbates, such as hydrogen or mixed hydrogen/hydroxyl groups.  The resulting removal of the SiC surface states from the band gap leads to a near-perfect restoration of the defect's emission frequency and its photostability by the elimination of different charge-conversion pathways.

 Although effective and experimentally straightforward to implement, the use of atomic and molecular adsorbates is limited by their thermal and chemical stability in the long-term~\cite{stability1,stability2,Shinohara2005}. Hence, a more robust passivation scheme is needed that retains the different benefits of using simple adsorbates, but eliminates their limitations.  That this can be achieved was shown in an experimental work by Polking \textit{et al.}~\cite{Polking_deLeon_SiC_surface2018}. The authors grew an epitaxial layer of wurtzite aluminum nitride (AlN) on a SiC surface and reported improved photostability of the positively-charged carbon antisite-vacancy defect complexes within SiC.  
These were promising experimental results, however, to date, there are no theoretical studies to explain how AlN passivation restores a defect's photostability, or its effects on other SiC defects.  For example, it is not known if such a passivation scheme will also improve the charge-state stability of the silicon monovacancy, which is bright only in the negatively charged state. In this first principles work, we study AlN-coverage of SiC surfaces by modeling a core-shell nanowire with a SiC core and an AlN shell. Using $\mathrm{V_{Si}^{-1}}$ in SiC as a proof-of-principle defect, we demonstrate that AlN effectively removes SiC surface states from the band gap, restoring the defect's optical properties. We also report the existence of novel silicon vacancy-based defects at the SiC-AlN interface with distinct spin and optical properties.

\begin{figure*}[htb]
  \centering
  \includegraphics[width=.75\textwidth]{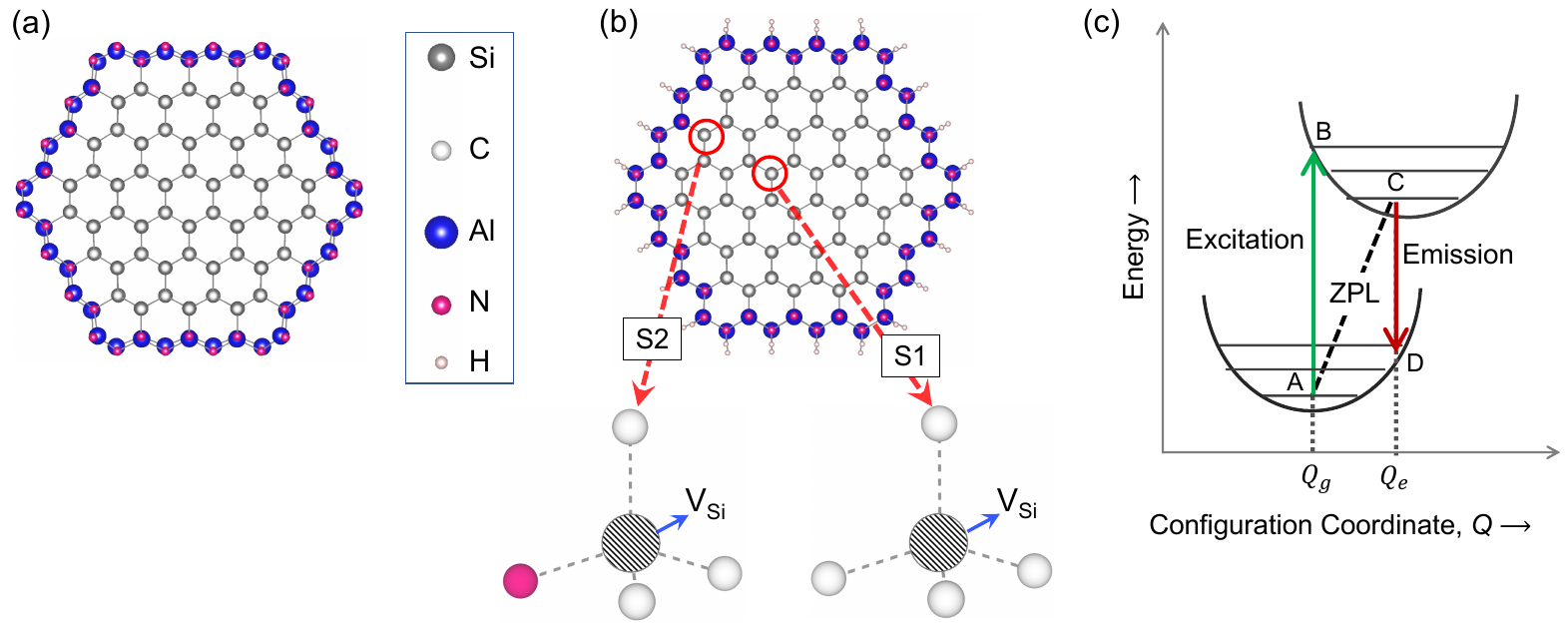}
   \caption{Aluminum nitride passivated silicon carbide nanowire (NW). (a) SiC/AlN NW (cross-sectional view), (b) Hydrogen (-H) terminated SiC/AlN NW to mimic a thicker AlN shell (cross-sectional view). Passivation of SiC core with AlN shell creates two distinct defect sites (S1 and S2) relative to the interface between SiC core and AlN shell. (c) Franck-Condon picture used for interpreting $\Delta$SCF results.}
   \vspace{-6pt}
     \label{Figure1}
   \end{figure*}

\section*{Computational Details}
Our density functional theory (DFT)-based spin-polarized calculations were performed using the QUANTUM ESPRESSO package~\cite{QE-2009,QE-2017}. %using a plane-wave basis set.  
We used ultrasoft pseudopotentials~\cite{USPP_vanderbilt1990}, which allow for low energy cut-offs of 50\,Ry for expanding wave functions and 400\,Ry for charge densities.  In addition, all of the results reported in this work are for structures that were relaxed until the forces on the atoms were smaller than $10^{-3}$\,Ry/a.u. We used the generalized gradient potential approximation (GGA)~\cite{GGA} of Perdew, Burke, and Ernzerhof (PBE)~\cite{PBE} to account for the exchange-correlation effects.  The PBE functional was chosen instead of the prohibitively expensive Heyd-Scuseria-Ernzerhof (HSE) hybrid functional~\cite{HSE03, HSE06} because:  (i) it made our calculations for large nanostructures feasible, and (ii) our emphasis is on properties that are expected to remain the same whether PBE or the computationally  expensive hybrid functionals are used. For example, in this work, we explored if AlN can effectively passivate SiC surfaces, and if this passivation scheme changes the optical properties of the monovacancies.  For the latter, we considered changes between the quantum emission frequencies of the monovacancies in the core-shell NW and bulk.  Since such a change in emission frequency is a difference of differences in total energies, any error in our predictions was minimized.   

%%%%%%%%%%%%%%%%%%%%%%%%--- structure description---%%%%%%%%%%%%%%%%%%%%%%%%%%%%%%%%%%%%%%%%%%%%%%%%%%%%%%%%%%%%%%%%%%%%%%%%%%%%%%%%%%%%%%%%%%%

Our structure is a core-shell NW with a 2H-SiC polytype at its core and a shell of wurtzite AlN.  As done in recent studies~\cite{Joshi2022,Sayou_PRMater2024}, we chose to use the 2H-SiC polytype for the NW because it ensured that the as-created stoichiometric nanostructures were non-magnetic, with no magnetic moment on any atom. This prevented any surface-related spin from interacting with the spin of $\mathrm{V_{Si}^{-1}}$, allowing us to concentrate on the role that the finite-size and surface effects play in modifying the defect's properties. 
Our calculations show the bulk lattice constants, $a$ (in-plane) and $c$ (out-of-plane) of wurtzite AlN and 2H-SiC differ by 1.3\% and  2.3\%, respectively. Hence, wurtzite AlN is a good choice for the passivation layer due to its small lattice-mismatch with 2H-SiC. Note that other hexagonal polytypes, such as 4H- and 6H-SiC, are the more commonly explored hosts of quantum emitters~\cite{Christle2015,Fuchs2015,falk2013polytype,widmann2015coherentRT,Seo2016,nagy2019highfidelity,Carter2015,Wolfowicz2020,Breev2022,CarterDev2025,Stuermer2025}. These hexagonal polytypes differ from each other and from 2H-SiC in the stacking of the tetrahedrally-bonded silicon and carbon atoms. The differences in stacking sequences result in different local crystal fields with distinct h- and k-sites, as well as different band gaps of the hexagonal polytypes. In turn, these differences result in variations in the excited state properties, such as the quantum emission frequencies, of their quantum emitters~\cite{falk2013polytype}. Nevertheless, qualitatively, the results from our work on $\mathrm{V_{Si}}$-based defects in 2H-SiC are general and should apply to other SiC polytypes as well. 
It should also be added that the feasibility and effectiveness of using AlN as a passivation layer for a different polytype (4H-SiC) has already been established by Polking \textit{et al.} ~\cite{Polking_deLeon_SiC_surface2018}.

The ideal unpassivated SiC/AlN NW, shown in Figure~\ref{Figure1}(a), has a  diameter of $21.8$\,\AA{}. The NW  is periodic along the $[0001]$ direction (c-axis), with a vacuum of about $13.5$\,\AA{} separating the images of the NW in the lateral direction (perpendicular to the c-axis), minimizing the interaction between the periodic images of the nanowire. We use a gamma-centered k-point grid of $1\times1\times6$, created according to the Monkhorst-Pack scheme~\cite{Monkhorst}.  The unpassivated core-shell NW (from here on, called SiC/AlN NW) has a total of 384 atoms, with a SiC core of $216$ atoms and an AlN shell consisting of $168$ atoms.  The surface of the as-created SiC/AlN NW has threefold- and four-fold coordinated atoms~\cite{Joshi2022}. The undercoordinated atoms on the surface of this NW were further passivated with hydrogen (H) atoms for reasons that will be discussed in the next section. It is this passivated core-shell structure (SiC/AlN-H NW) that was used in our study. Figure~\ref{Figure1}(b) shows the cross-sectional view of the SiC/AlN-H NW with a total of 480 atoms. Further structural details are provided in Supplementary Figures S1(a) and (b) in the Supporting Information (SI). There are two distinct possibilities for the placement of silicon monovacancies relative to the interface between SiC and AlN: within the SiC core (i.e away from the interface) or at the interface itself. Two such sites -- S1 and S2 -- are shown in Figure~\ref{Figure1}(b). In this work, we used silicon vacancies at these sites to determine the efficacy of AlN in removing dangling bonds and restoring the optical properties of the monovacancies. Since the hydrogen-passivated SiC NW used in the older work by Sayou \textit{et al.}~\cite{Sayou_PRMater2024} had a smaller diameter, we performed additional calculations for a hydrogen-passivated SiC NW (henceforth called SiC-H NW), consisting of 480 atoms for sake of direct comparison with the results for defects in the SiC/AlN-H NW. Supplementary Figures S1(c) shows the cross-sectional (top) view of the SiC-H NW.

\begin{figure*}[htb]
  \centering
  \includegraphics[width=.9\textwidth]{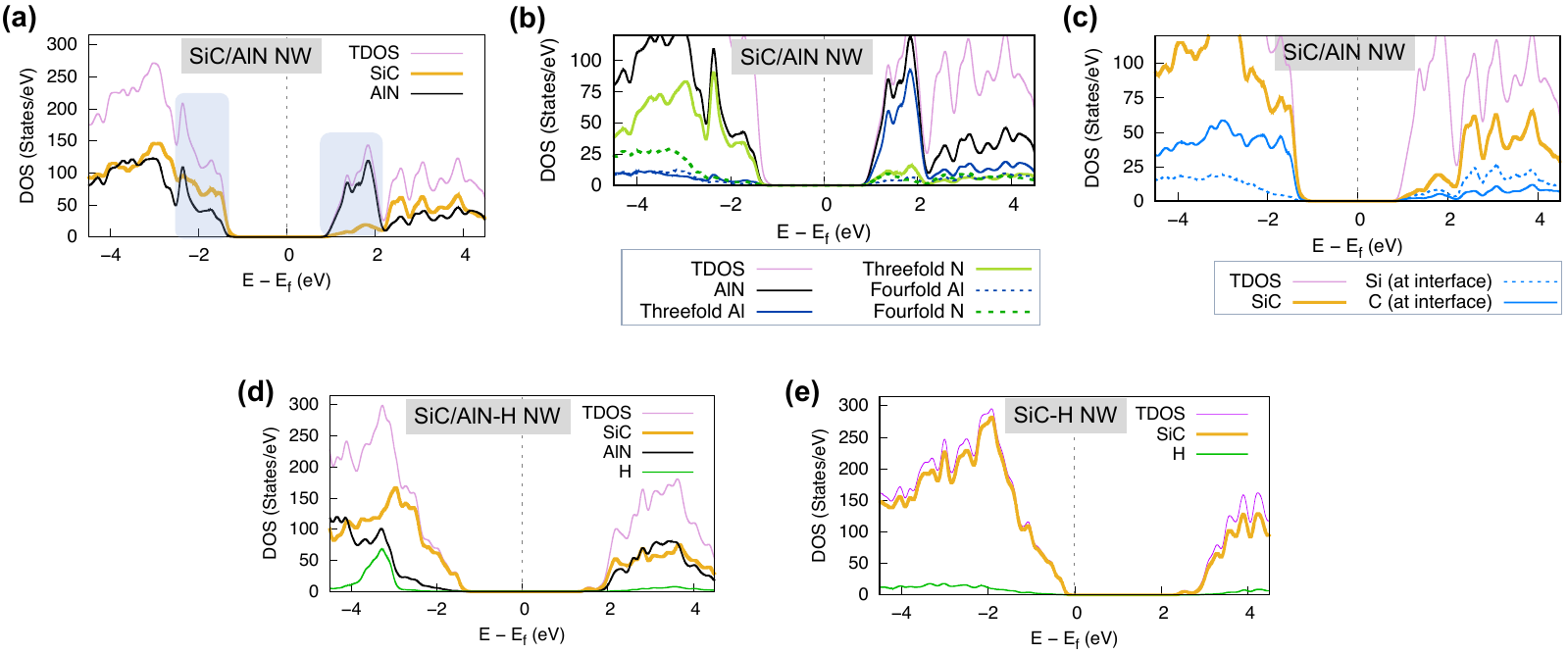}
   \caption{Electronic structure properties of defect-free 2H-SiC NWs. (a) Total and projected density of states (DOS) for unpassivated/bare SiC/AlN NW, showing the narrowing of the bandgap in the core-shell structure to 2.17\,eV due to surface and interface states (highlighted in blue).  (b) DOS projected onto the threefold and fourfold coordinated Al and N atoms in the AlN shell, showing their contribution to the states at the band edges of the bare core-shell NW. (c) DOS projected onto the Si and C atoms at the interface between the core-shell NW, showing contributions from the interfacial carbons to the DOS at the valence band edge. (d) The surface and interface states are removed by hydrogenation of the core-shell NW.  (e) DOS for SiC-H NW for comparison.}
      \vspace{-6pt}
     \label{Figure2}
   \end{figure*}

In order to study the optical properties of the near-surface $\mathrm{V_{Si}^{-1}}$ in SiC/AlN-H NW, we obtained the zero-phonon line (ZPL) using the $\Delta$SCF method~\cite{KaxirasNV}, which has been successfully employed to study excitations between spatially-localized defect states of deep-level defects in different wide band gap semiconductors~\cite{PDEV_hBN_2020,Joshi2022,narayanan2023,Sayou_PRMater2024}.  In the $\Delta$SCF method, the occupation of the defect states is constrained to mimic the photo-excitation process as described within the Franck-Condon model shown in  Figure~\ref{Figure1}(c).  Point A in the Franck-Condon picture represents the system in the electronic ground state with the ground-state equilibrium ionic positions (generalized coordinate $Q_{g}$). The spin-preserving, vertical optical excitation, $\mathrm{A} \rightarrow \mathrm{B}$, involves emptying of a filled defect state and the simultaneous occupation of the previously empty defect state (without structural relaxation). Point C in the Franck-Condon picture is accessed by allowing for the ionic relaxation (excited-state generalized coordinate $Q_{e}$) that follows the new electronic configuration in the excited-state.  Point D corresponds to the system with excited-state ionic positions, $Q_{e}$, and ground state electronic configuration. The total energy differences for the system at points $A$, $B$, $C$ and $D$ yield the following: (i) vertical transition upon absorption of a photon ($\mathrm{A} \rightarrow \mathrm{B}$), (ii) purely electronic ZPL transition ($\mathrm{A} \leftrightarrow \mathrm{C}$), (iii) the Stokes shift ($\mathrm{B} \rightarrow \mathrm{C}$) and (iv) the vertical de-excitation upon emission of a photon ($\mathrm{C} \rightarrow \mathrm{D}$).

\section{Results and Discussion}
 \subsection{Defect-free Core-Shell NW}

 The spatial and energetic proximity of the surface states to the defect states allows them to interfere with the photostability of the quantum emitters~\cite{Joshi2022,Sayou_PRMater2024}.  This can be avoided by passivating SiC dangling bonds within the passivation-scheme of choice, be it simple adsorbates or a coverage with another semiconductor. In principle, with the AlN-passivation of SiC one should obtain an electronic structure where the surface-states from the AlN shell are far removed from defect states (at least spatially). However, as can be seen in the density of states (DOS) plot in Fig~\ref{Figure2}(a) for the bare/unpassivated SiC/AlN NW, we have surface states in the band gap. Figures~\ref{Figure2}(b) and (c) plot the DOS projected onto the $2s$ and $2p$ orbitals of atoms belonging to the AlN shell and the interfacial Si and C atoms of the SiC core. A closer look at the two figures shows that it is the threefold-coordinated Al and N atoms, along with the C atoms at the interface that contribute the most to the states at the band edges of the core-shell NW.  One expects to find the surface states originating from threefold-coordinated Al and N atoms at the band edges~\cite{Sayou_PRMater2024}. However, the reasons for the presence of the states from the interfacial C atoms at the band edges are not self-evident. For this, we need to consider the skin-bond contraction effect, which is found in very small nanostructures~\cite{Huang2008SkinEffect}. The skin contraction occurs due to the strengthening of the bonds at the undercoordinated surface sites, which aids in a reduction of the surface energy.  The outcome of the bond-contraction is evident in Figure~\ref{Figure1}(a), which shows a marked change of hybridization from the ideal $sp^{3}$ to a mixed $sp^{2}$-$sp^{3}$ type for the surface atoms [compare with Figure~\ref{Figure1}(b)]. This change in hybridization at the surface can also be seen in the Supplementary Figure~S2(a).  These structural changes at the surface propagate to the interior of the NW~\cite{Joshi2022,Sayou_PRMater2024}. Since we have a thinner shell, these structural changes at the AlN surface propagate to the SiC-AlN interface, affecting the structure of SiC at the interface. 
 The presence of surface states and the states from the interfacial atoms, in turn, reduces the band gap of the SiC/AlN NW to 2.17\,eV, which is smaller than the theoretical band gap (2.31\,eV) of bulk 2H-SiC. The calculated electronic structure of the SiC/AlN NW are an artifact of our core-shell model, in which the AlN shell is extremely thin. However, in order to demonstrate that AlN grown on SiC is an effective passivation scheme for improving optical properties of near-surface defects, it should be shown that the hybridization between the detrimental surface states and the defect states from $\mathrm{V_{Si}^{-1}}$ in SiC is minimal. This can be achieved only if the defect states are energetically and/or spatially far removed from surface states.  In other words, we need to either create a structure with much thicker shell or a structure that mimics the behavior of a thicker shell.  Due to the prohibitive cost of calculations involving thicker shells, we decided to emulate a much thicker AlN shell by passivating the structure with hydrogen.

  In order to mimic the  behavior of a much thicker AlN shell, we carefully considered if the undercoordinated atoms on the surface should be passivated using real hydrogens with a valence charge of $1\,e$, or pseudohydrogens (PH) with fractional charges~\cite{Shiraishi_1990,SuHuaiWei_PH_PRB,Yoo2021} of $1.25\,e$ and $0.75\,e$ for passivating the threefold coordinated  Al and N atoms, respectively.  
 We found that the properties of the core-shell NWs that influence the defect states -- band gaps and the DOS around the band edges  --  remained the same whether we used real H or PH atoms [see Supplementary Figure  S3]. We attribute this to the straddling of the smaller band gap of SiC by AlN's band gap~\cite{Choi_band_alignment2005} as discussed in detail within Supplementary Note I. In addition, whether we use real H or PH atoms, the surface passivation restores the $sp^{3}$ character of bonds  between the atoms at the surface [see Supplementary Figure~S2 (b)]. For the sake of completeness, we present our results for real H-passivation in the main text, while those for the PH-passivated core-shell structures are presented in the supporting information.  A full hydrogen coverage of all threefold coordinated surface atoms (totaling 96 atoms) removed the detrimental surface and interfacial states from the band edges as seen in Figure~\ref{Figure2}(d).  As a consequence, the band gap of the resulting SiC/AlN-H NW is 2.85\,eV.  As expected, this is larger than the band gap for the bulk material due to quantum confinement effects that come into play in a nanostructure.  For comparison, we also plot the DOS of the SiC-H NW, which shows a band gap of 2.77\,eV.  

\begin{figure*}[htb]
    \centering
    \includegraphics[width = 0.90\textwidth]{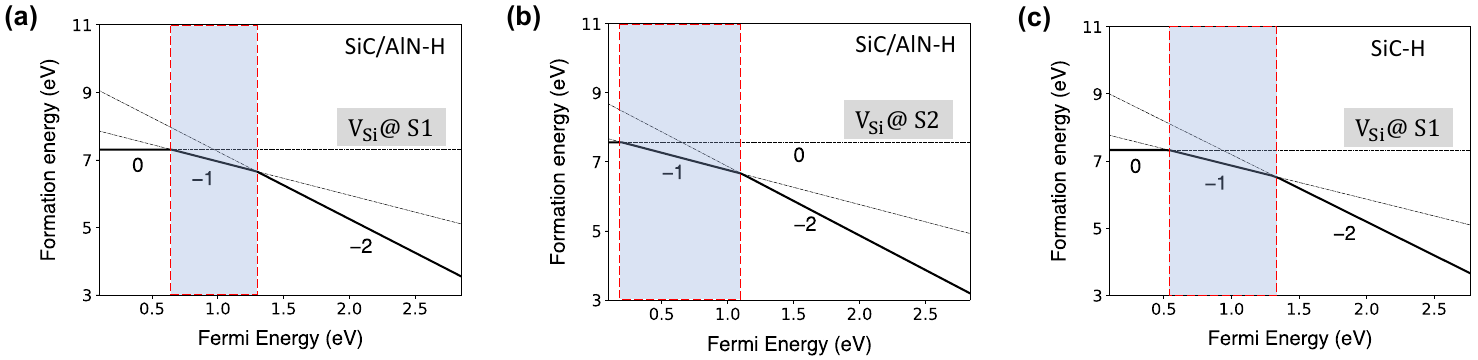}
    \vspace{-6pt}
    \caption{Formation energies ($\Delta E_{form}$) for $\mathrm{V_{Si}}$ in different charged states as a function of the Fermi energy level (electronic chemical potential) in real hydrogen passivated NWs with the monovacancy at the (a) S1 (interior) site within the core-shell SiC/AlN-H NW, (b) S2 (interface) site of the SiC/AlN-H NW, and (c) interior site of the SiC-H NW.  In each plot, the Fermi energy level is given with respect to the valence band maximum.  The formation energies are calculated assuming carbon-rich conditions. The favorable doping conditions at which the negatively charged state ($\mathrm{q}=-1$) of the defect is stable, are highlighted. %The silicon-rich conditions  yield similar plots (not shown), but with slightly higher formation energies.
    }
       \vspace{-6pt}
    \label{Figure3}
\end{figure*}

\subsection{Silicon vacancy in passivated SiC/AlN-H NW}

 As discussed in the previous section, the hydrogenation of the surface of our extremely thin AlN-shell allowed us to emulate the behavior of experimentally-relevant thicknesses of AlN shells~\cite{Polking_deLeon_SiC_surface2018}. In doing so, we also remedied the excessive structural changes at the SiC-AlN interface, which are artifacts of the skin-contraction effects and not due to lattice mismatch. 
Our next step was to study properties of $\mathrm{V_{Si}^{-1}}$ at the S1 and S2 sites in SiC/AlN-H. Before presenting our results for $\mathrm{V_{Si}^{-1}}$ at these sites, it is instructive to see how the presence of the AlN shell may change the charge state stability.
To do so, we calculated the defect formation energies for $\mathrm{V_{Si}}$ at the S1 (interior) and S2 (interface) sites within the SiC/AlN-H NW. The Freysoldt-Neugebauer-Van de Walle (FNV) correction scheme~\cite{FNV_2009} as implemented in the CoFFEE code~\cite{coffee_2018} was used to correct for the spurious Coulomb interactions between the charged defect and its periodic images.  

Figures~\ref{Figure3} (a) and (b) are plots of defect formation energies vs electronic chemical potential (Fermi energy) referenced to the valence band maximum for the SiC/AlN-H NW.  For comparison, we have also provided the formation energies  of the defect (at the interior site) in SiC-H NW in Figure~\ref{Figure3} (c).  All results are obtained under carbon-rich (i.e. silicon-poor) conditions, which would be the growth condition that favors formation of silicon vacancies.  The highlighted regions reveal that the bright, negatively-charged state ($\mathrm{q}=-1$) of $\mathrm{V_{Si}}$ is stable over a considerable range of n-doping within the SiC/AlN-H and SiC-H NWs.  Also note that the charge transition levels [(0/1-) and (1-/2-)] for the defect at the S2-site are located lower in the band gap as compared to those for the S1-site defects in the SiC/AlN-H and SiC-H NWs [see Figs~\ref{Figure3} (a) and (c)]. Although the formation energies for monovacancies in SiC/AlN-PH follow similar qualitative trends [see Supplementary Figure S4 and Supplementary Note II], the $\mathrm{q}=-1$ charge state for the defect at the S2-site is stable over a narrower range of chemical potential in SiC/AlN-PH than in SiC/AlN-H.

\begin{figure*}[htb]
  \centering
  \includegraphics[width=.9\textwidth]{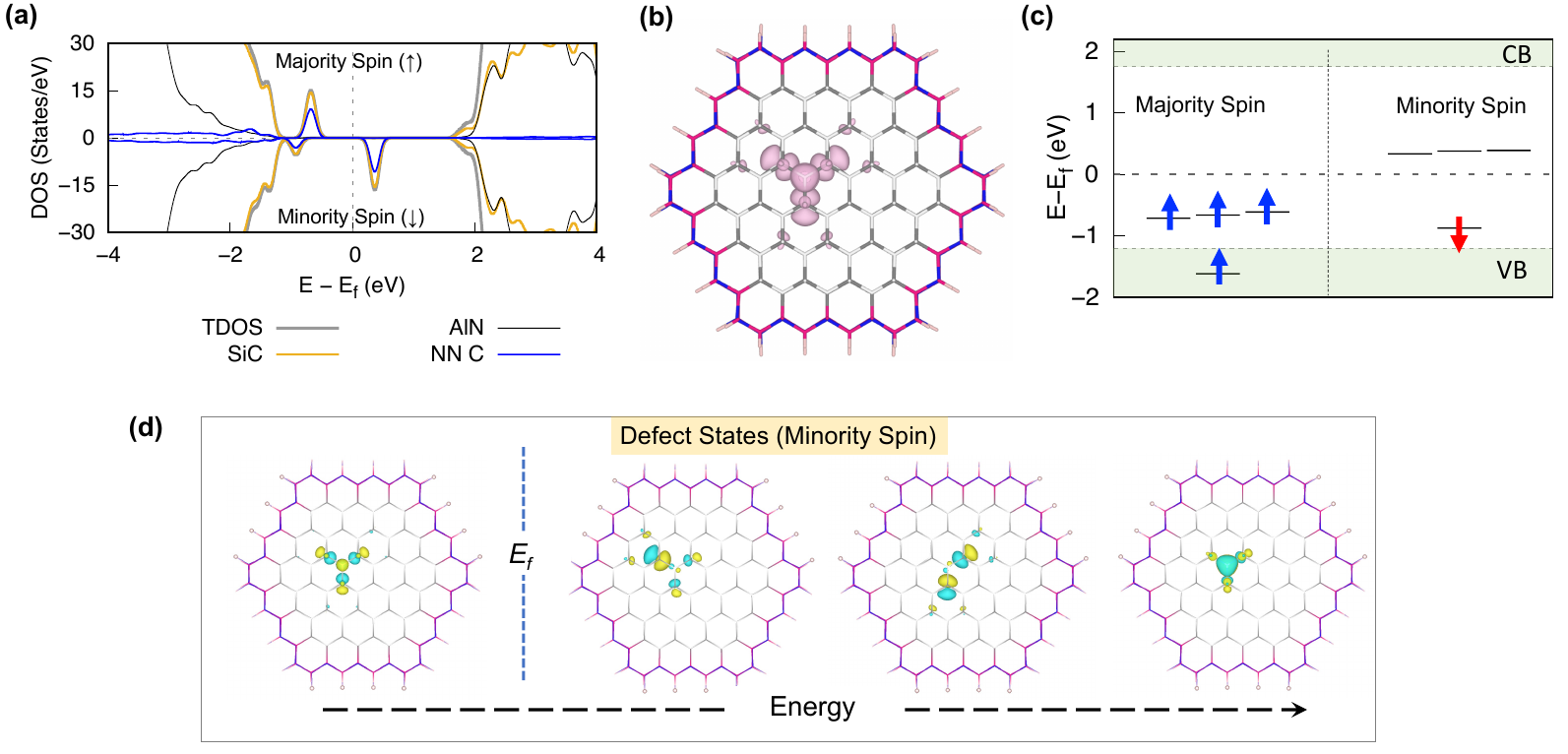}
   \caption{SiC/AlN-H NW with a $\mathrm{V_{Si}^{-1}}$ defect in the interior (S1 site). (a) Total density of states (TDOS) in thick gray line, along with the DOS contributed by the SiC core (in gold), AlN shell (thin black line) and the four NN carbon-atoms (in blue).  (b) Spin density [$\Delta \rho^{spin}$] isosurface plot for the monovacancy at the S1 site, showing that the most of the spin-3/2 of the defect comes from the unpaired electrons in the $sp^{3}$-hybridized $2s$ and $2p$ orbitals of the carbons surrounding the defect.  Here, $\Delta \rho^{spin} = \rho^{\uparrow}-\rho^{\downarrow}$ is the difference in the spin-up and spin-down charge densities.  (c) Energy-level diagram, showing the positions of the defect states in majority spin (spin-up) and minority spin (spin down) channels at the $\Gamma$-point.  The optically-active defect states in the minority spin channel are well-separated from the valence band (VB) and conduction band (CB) states.  
 (d) The charge density plots of the optically-active empty and filled state for SiC/AlN-H NW at the $\Gamma$-point.  Yellow (blue) color corresponds to positive (negative) isovalues.}
   \vspace{-6pt}
     \label{Figure4}
   \end{figure*}

\subsubsection{$\mathrm{V_{Si}^{-1}}$ at the S1 (interior) site in SiC/AlN-H NW}

The singly charged $\mathrm{V_{Si}^{-1}}$ defect at the interior site, S1, has a total of five electrons in the dangling bonds left behind by the monovacancy. It is a spin-3/2 defect, just as in bulk SiC~\cite{Joshi2022,Carter2015,soykal2016silicon,economou2016spin}. 
Figures ~\ref{Figure4}(a)-(d) give the details of the electronic structure for this defect. Figure~\ref{Figure4}(a) is a plot of TDOS (in gray) of the defective SiC/AlN-H NW, along with the projected DOS contributed by the SiC core (in gold), AlN shell (in thin black line) and the four NN carbon-atoms, $\mathrm{C_{NN}}$'s surrounding the defect (in blue). The plot clearly shows that the defect creates localized defect states in the band gap. The DOS contributed by SiC shows that these defect states are contributed by the core with negligible contribution from the AlN shell. The DOS contribution from the four $\mathrm{C_{NN}}$'s and the isosurface plot for the spin density [$\Delta \rho^{spin} = \rho^{\uparrow}-\rho^{\downarrow}$] in Figure~\ref{Figure4}(b) further shows that the most of the spin-3/2 of the defect comes from the unpaired electrons in the dangling bonds associated with the carbon atoms surrounding the defect.  The localized nature of the defect states itself can be traced back to the highly localized nature of carbon's $sp^{3}$-hybridized $2s$ and $2p$ orbitals that form the dangling bonds. The large exchange interactions between the unpaired electrons in the nearly-dispersionless defect states results in the spin-splitting of the defect states seen in Figure~\ref{Figure4}(a)~\cite{Dev_PRL_DeepDefects_2008,Dev_PRB_DeepDefects_2010,Dev_PRB_NW_2010}.

To understand the origin of spin-3/2 for the defect, we show the energy-level diagram in Figure~\ref{Figure4}(c). The defective NW structure has the $C_s$ symmetry, unlike the $C_{3v}$ symmetry for the defect in the bulk~\cite{Joshi2022,Sayou_PRMater2024}, and thus, all of the defect states are singlets.  The five electrons in the dangling bonds fill the ``atomic-like" localized defect states such that there are three more electrons in the spin-up channel (majority-spin) than the spin-down channel (minority spin), resulting in the spin-3/2 for the defect.   The PH-passivated SiC/AlN NW gives identical results for the defect at the S1-site  [see Supplementary Figure S5].

\begin{figure*}[ht]
  \centering
  \includegraphics[width=.8\textwidth]{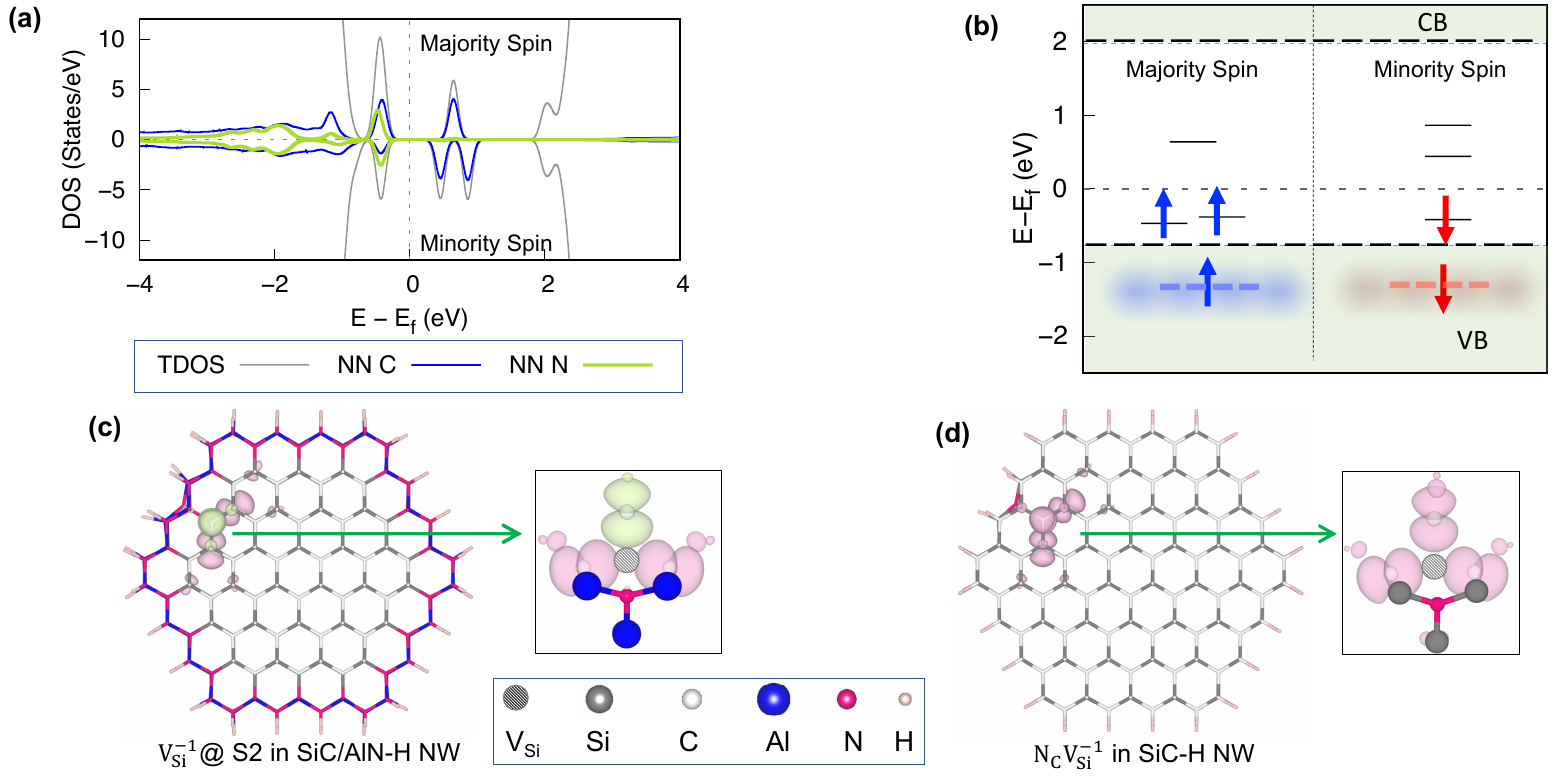}
   \caption{SiC/AlN-H NW with a $\mathrm{V_{Si}^{-1}}$ defect at the core-shell interface  (S2 site). (a) Total density of states (TDOS), along with the DOS contributed by the nearest neighboring (NN) atoms surrounding the defect.  The latter include contributions from the $2s$ and $2p$ orbitals of the three $\mathrm{C_{NN}}$ (in blue) and $\mathrm{N_{NN}}$ (in green).  (b) Energy-level diagram, showing the positions of the in-gap defect states in majority spin (spin-up) and minority spin (spin down) channels at the $\Gamma$-point.  The lowest defect states, which are resonant with the valence band (VB), mix with the VB states and can no longer be uniquely identified. They are included for the sake of completeness.   
 (c) Spin density plot for the monovacancy at the S2 site. Pink (green) color corresponds to positive (negative) isovalues.  Also shown is a close-up of the spin density plot (side-view), consisting of the three $\mathrm{C_{NN}}$'s and the $\mathrm{N_{NN}}$-atom, along with the Al atoms bonded to $\mathrm{N_{NN}}$. %The magnetic moment on the axial $\mathrm{C_{NN}}$ is antiferromagnetically aligned to the moments on the two basal $\mathrm{C_{NN}}$'s, resulting in a net spin-1/2 for the defect. 
 (d)  Spin density isosurface plot for the negatively charged $\mathrm{N_{C}V_{Si}}$-center in the H-SiC NW [at the same site as the defect in (c)]. The close-up (side-view) shows $\mathrm{C_{NN}}$'s and the $\mathrm{N_{NN}}$-atom, along with the Si atoms bonded to $\mathrm{N_{NN}}$.}%highlighting the bonding and/or antibonding character of the defect states} %The charge density plots of the optically-active empty and filled state for SiC/AlN-H NW at the $\Gamma$-point.  Yellow (blue) color corresponds to positive (negative) isovalues.}%highlighting the bonding and/or antibonding character of the defect states.
    \vspace{-6pt}
     \label{Figure5}
   \end{figure*}

The optical excitation happens between the filled and the empty defect states in the spin-down channel. Figure~\ref{Figure4}(d) shows the charge density plots for these optically-active empty and filled states.  The calculated value of the ZPL is 1.12\,eV, which differs from the bulk ZPL-value of 1.19\,eV by about 75\,meV. As discussed in earlier works~\cite{Joshi2022,Sayou_PRMater2024}, the difference in ZPL for a defect close to the surfaces and/or interfaces can be attributed to several inter-connected factors, with the most significant being: (i) local strains, (ii) differences in Stokes shift, which result from different structural relaxation around the defect upon photoexcitation, and (iii) different placements and hence gaps between the optically-active defect states, resulting in different absorption energies.   To estimate the strain-induced change in the ZPL, we used the experimentally-determined strain-coupling parameter, 1130\,meV(strain$)^{-1}$, for strain in the basal direction~\cite{nanostr_VSi_expt_2020}.  The bond lengths along the basal plane at the S1-site are  0.39\% longer as compared to the ideal basal distance. This yields a ZPL-change of 4.38\,meV, showing that strain is relatively less important in changing  the ZPL.  The Stokes shift, given by $\mathrm{E_{B}-E_{C}}$ in Figure~\ref{Figure1}(c), is calculated to be 122.3\,meV for the defect at S1, while it is 104.9\,meV for the defect in the bulk material.  The 17.4\,meV difference in the Stokes shift accounts for some of the calculated ZPL shift. In the case of $\mathrm{V_{Si}^{-1}}$ at the S1-site in the SiC/AlN-H NW,  the largest contributor to the ZPL-shift is from the differences in vertical absorption energies [$\mathrm{E_{B}-E_{A}}$ in Figure~\ref{Figure1}(c)] for the defect at the S1 site (1.24\,eV) and for the bulk (1.30\,eV), with a 57.89\,meV decrease in the calculated value of $\mathrm{E_{B}-E_{A}}$ for the defect in core-shell NW as compared to the bulk.  To see the influences of the AlN-shell on the optical properties of the interior defect, we also calculated the optical properties of $\mathrm{V_{Si}^{-1}}$ at an interior-site in the SiC-H NW.  We find that the ZPL for this defect (1.16\,eV) is much closer to that for the bulk SiC, with a 30\,meV difference in the ZPL. For the monovacancy in the interior site of SiC-H NW,  the contributions to the ZPL-shift from strain, changes in Stokes shift and the vertical excitation are found to be 1.57\,meV,  11.77\,meV and  16.87\,meV, respectively. Hence, the largest change produced by AlN shell is in the placement of the defect states within the band gap (as compared to a SiC-H NW), owing to the changes in the electronic structure properties upon creating a core-shell NW.

%Spin density plot for the monovacancy at the S2 site, showing that most of the spin-1/2 of the defect is contributed by NN carbon-atoms, with the spin on the NN nitrogen being nominally zero. 
\subsubsection{$\mathrm{V_{Si}^{-1}}$ at the S2 (interface) site in SiC/AlN-H NW}

At the S2 site, a negatively charged silicon vacancy is surrounded by three nearest neighbor carbons and one nitrogen atom from the shell [see Figure~\ref{Figure1}(b)]. One might na{\"i}vely expect this defect to behave like a negatively charged NV-center in SiC, wherein a substitutional nitrogen replaces one of the nearest neighbor carbons of the silicon monovacancy. In  bulk 2H-SiC, such an NV center, with an N-atom substituting a basal C-atoms, is a spin-1 defect. However, for the negatively charged defect at S2 we found two solutions -- a spin-1/2 (low-spin) ground state solution and a spin-3/2 (high-spin) metastable solution. The latter was higher in energy as compared to the low-spin state solution by 163.30\,meV. Interestingly, neither of the two solutions can be explained by assuming that the defect at the S2-site has a total of six electrons in four dangling bonds surrounding the monovacancy.  In fact, a singly negatively-charged defect behaves as if it has a total of five electrons in the defect states. In what follows, we will concentrate on the lower-energy spin-1/2 solution for the interfacial defect, but similar arguments apply to the metastable solution as well. 
\begin{figure*}[ht]
  \centering
  \includegraphics[width=.75\textwidth]{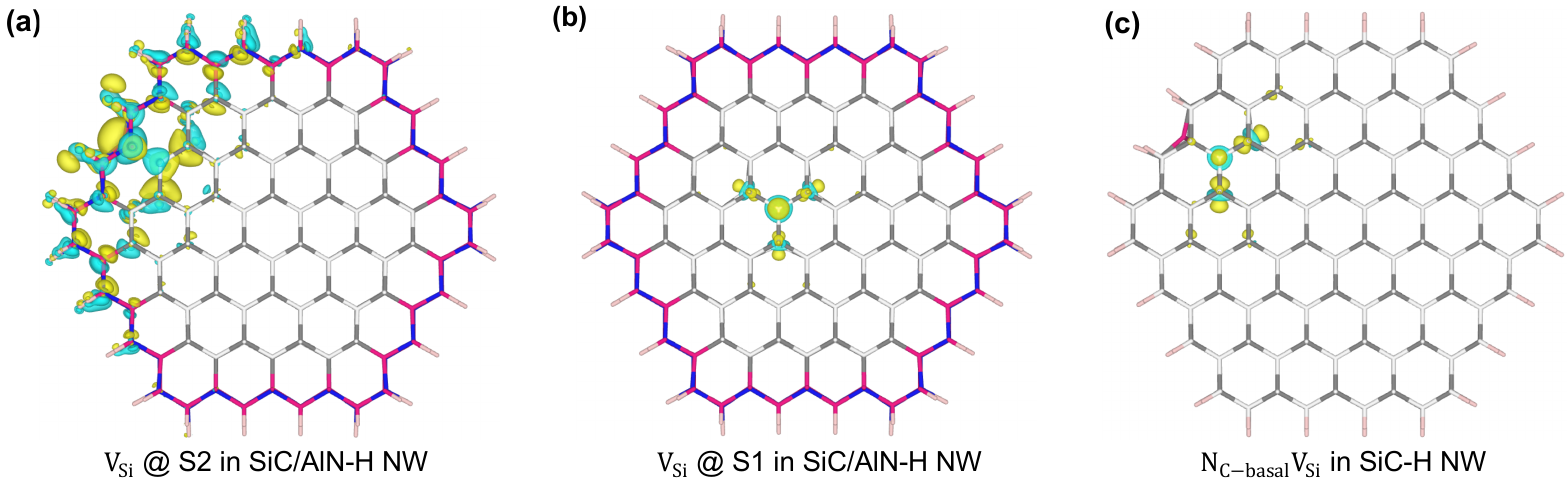}
   \caption{Charge density difference [$\Delta \rho =  \rho^{q=-1}-\rho^{q=0}$] for: (a) $\mathrm{V_{Si}}$ at the S2 (interface) site in the SiC/AlN-H NW, showing how the additional charge gets delocalized over a large region, including parts of the AlN shell and SiC core, (b) $\mathrm{V_{Si}}$ at the S1 (interior) site in the SiC/AlN-H NW, and (c) an NV center in the SiC-H NW, with the last two cases showing the localized nature of the additional charge. %In all cases, for the neutral state of the defects, the equilibrium ionic positions of the respective charged defects are used. 
 Yellow (blue) color corresponds to regions with charge accumulation (depletion). %with the overall picture showing the sense in which charge is redistributed upon addition of an electron to the respective neutral defects.
 }
    \vspace{-6pt}
     \label{Figure6}
   \end{figure*}

We first concentrate on explaining the calculated electronic structure to explain the spin-1/2 of the defect and then tackle the question: `` where is the sixth electron?''   In Figure~\ref{Figure5}(a), we plot the total and the partial DOS plots for the spin-1/2 defect at the interface, showing that the defect introduces very localized (sharp) defect states in the band gap. The DOS contributions from the $2s$ and $2p$ orbitals of the NN carbons atoms (in blue) and nitrogen atom (in green) show that the in-gap sharp states originate in the dangling bonds on the surrounding atoms. The energy-level diagram is shown in  Figure~\ref{Figure5}(b). The defect states resonant with VB are shown as ``fuzzy'' lines as these defect states hybridize with the states in the VB and their position can no longer be uniquely given.  A comparison with the energy level diagram for the interior defect [Figure~\ref{Figure4}(c)] shows that the three nearly-degenerate defect states of the interior defect have undergone  larger crystal field splitting at the interface, with the distribution of the five electrons resulting in net spin-1/2 for the interfacial defect.  We obtain a qualitatively similar energy level diagram for the S2-site defect in the SiC/AlN-PH NW  [see Supplementary Figure S6 and Supplementary Note-II]. 

 The spin density for the defect at the interface is shown in Figure~\ref{Figure5}(c). Pink (green) color corresponds to positive (negative) isovalues. The magnetic moment of the axial NN carbon is antiferromagnetic aligned to the moments of the NN carbons along the basal plane, resulting in the net spin-1/2 for the defect.  The close-up of the spin density plot (side-view) shows only the three $\mathrm{C_{NN}}$ atoms and $\mathrm{N_{NN}}$ atom, along with the Al atoms bonded to the $\mathrm{N_{NN}}$-atom.  The three Al atoms are shown to highlight that, although nominally a negatively-charged NV center, this defect has a different chemical environment as compared to the same defect in SiC bulk or SiC NW without an AlN shell. To demonstrate that the antiferromagnetic alignment is not a result of proximity to the surface, we created a negatively charged $\mathrm{N_{basal-C}V_{Si}}$ defect center (subsequently referred to as the ``NV-center'') in a SiC-H NW at the same site. We find that the near-surface, negatively-charged NV center in SiC-H NW is a spin-1 defect (similar to the NV-center in the bulk SiC) with a spin-density distribution shown in Figure~\ref{Figure5}(d).  Hence, neither the nanostructuring of SiC, nor the proximity to the surface are responsible for the behavior of the interfacial defect in SiC/AlN-H NW.

%In order to determine why the singly-charged defect appears to have only five electrons,
In order to answer the question posed in the previous paragraph (``where is the sixth electron?"), we calculated the charge-density differences between the singly-charged and neutral defects [$\Delta \rho =  \rho^{q=-1}-\rho^{q=0}$] for three systems: $\mathrm{V_{Si}}$ at the S2 (interface) site in SiC/AlN-H NW (effectively, an NV center),  $\mathrm{V_{Si}}$ at the S1 (interior) site in SiC/AlN-H NW, and an NV center in a SiC-H NW.  The charge densities for the neutral charge states of all three defects were calculated with the respective equilibrium geometries obtained for the three negatively-charged structures.  Figures~\ref{Figure6}(a)-(c) are the plots of the charge density differences for the three systems. Yellow (blue) color corresponds to regions with charge accumulation (depletion). Overall, Figs~\ref{Figure6}(a)-(c) show the sense in which charge is redistributed upon addition of an electron to the respective neutral defects. The charge density differences for the monovacancy at S2 in Figure~\ref{Figure6}(a) shows that the electron added to the neutral defect at S2 does not get localized at the defect site. It is instead distributed over a large region, including the AlN shell.  One may suspect that this is either due to the nanostructuring of SiC itself or due to the proximity of the defect to the surface.   That this is not the case can be seen in the highly-localized charge redistribution upon the addition of an electron for the other two structures shown in Figs~\ref{Figure6}(b) and (c).  Hence, we find that the defect at S2 is fundamentally different from the other two defects and we attribute this to the different chemical environment of the defect at the AlN-SiC interface.  Interestingly, the interfacial defect acquires a spin-1 ground state when it is doubly-negatively charged. This charge state becomes stable for the $\mathrm{V_{Si}}$ at the S2 site for highly negatively doped NW as seen in Figure~\ref{Figure3}(b).  In our nanostructure, both $q=-1$ and $q=-2$ charge states of the interfacial defect have a spin-active ground state, with the magnetic structures being more stable than the non-magnetic structures by 387.03\,meV and 415.29\,meV, respectively.  The ZPLs for the singly and doubly charged monovacancies at the S2-site are 0.68\,eV and 0.77\,eV, respectively.   The latter is comparable to the ZPL of 0.84\,eV for the spin-1 $\mathrm{NV^{-1}}$-center (with N substituting a basal carbon) in bulk 2H-SiC. The difference in the ZPLs of the two spin-1 defects originates mostly from the much larger Stokes shift for the doubly-charged interfacial monovacancy (158.48\,meV) as compared to that for the singly-charged $\mathrm{NV}$-center in bulk 2H-SIC (82.23\,meV).

\section*{Conclusions}

The loss of photostability of near-surface, defect-based quantum emitters within nanostructured semiconductors is a severe bottleneck for their successful deployment in different quantum technologies.  %As proposed in an experimental work by Polking \textit{et al.}~\cite{Polking_deLeon_SiC_surface2018}, 
In this work, we have demonstrated that the core-shell structure, which consists of a SiC core passivated with an AlN shell, can effectively remove the detrimental surface states from the band gap of the nanostructured SiC host, elucidating the findings of an earlier experimental work~\cite{Polking_deLeon_SiC_surface2018}. In such a core-shell model, the AlN shell replaces simple atomic/molecular adsorbates, which will otherwise be used to passivate the surface states but are limited by their lack of long-term chemical and thermal stability. The presence of an AlN shell not only resolves the issue of charge-state instability of the near-surface $\mathrm{V_{Si}^{-1}}$ defect, it also restores the optical properties of this spin-3/2 defect. In addition, we identify a $\mathrm{V_{Si}}$-related defect at the interface of the SiC core and the AlN shell. The interfacial defect is nominally similar to the NV-center in SiC (with nitrogen along the basal plane), but displays different spin and optical properties from an NV-center in SiC due to its distinct chemical environment at the interface. Since this defect exists at the interface where other factors, such as strains and other defect complexes may be present, further careful investigations are needed to determine its usefulness under different conditions. Our atomistic study of defects in the SiC/AlN-H NW shows a promising means of resolving the optical stability issues of quantum emitters in nanostructured host semiconductors via creation of core-shell structures or growth/deposition of an appropriate semiconductor as a passivation layer on thin films of host semiconductors.  Lastly, as noted in our previous work~\cite{Sayou_PRMater2024}, although we have employed a small-diameter nanowire in our calculations,  the primary findings of this study are robust and should be applicable to the near-surface defects in larger nanostructures.

\section{Associated Content}
\noindent \textbf{Data Availability Statement}\\
The crystal structures data that support the findings of this study are available
from the corresponding author upon reasonable request.

%The data that support the findings of this study are available from the corresponding author upon reasonable request.
\noindent \textbf{Supporting Information}\\
\noindent The Supporting Information is available free of charge.
\noindent Supplementary figures and notes contrasting the resulting density of states, defect formation energies and the electronic structure of the defects in core shell nanowires, which is passivated with either real or  pseudohydrogens. 

\section{Author Information}
\noindent \textbf{Corresponding Author}\\

\noindent \textbf{Pratibha Dev} -- Department of Physics and Astronomy, Howard University, Washington, D.C. 20059, USA, Laboratory for Physical Sciences, College Park, Maryland 20740, United States;  orcid.org/0000-0002-6884-6737;
Email: pdev@lps.umd.edu\\

\noindent \textbf{Authors}\\
\noindent \textbf{Cyrille Armel Sayou Ngomsi} -- Department of Physics and Astronomy, Howard University, Washington, D.C. 20059, USA, Email: sayarmelo@gmail.com\\
\noindent \textbf{Sai Krishna Narayanan} --  Department of Physics, University of Maryland, College Park, Maryland 20742, USA, Email: saik98@umd.edu\\

%\noindent \textbf{Present Addresses}\\
%\noindent \textbf{Sai Krishna Narayanan} -- Department of Physics, University of Maryland, College Park, Maryland 20742, USA; Email: saik98@umd.edu\\

\noindent \textbf{Author Contributions}\\
PD conceived and directed the study. CASN and PD performed the DFT-based calculations. SKN calculated the defect formation energies with supercell-size corrections.  PD wrote the final manuscript with inputs from CASN and SKN.  All authors proof-read and reviewed the final revision of the manuscript.

\noindent \textbf{Notes}\\
The authors declare no competing financial interest.

\vspace{-0.1in}

\begin{acknowledgement}
The research performed at the Howard University was supported by the National Science Foundation under NSF Grants Nos DMR-1738076 and OAC-2118099. This work used the Expanse and Bridges2 clusters at SDSC and PSC, respectively, through allocation PHY180014 from the Advanced Cyberinfrastructure Coordination Ecosystem: Services \& Support (ACCESS) program, which is supported by National Science Foundation grants No. 2138259, No. 2138286, No. 2138307, No. 2137603, and No. 2138296.

\end{acknowledgement}

%\bibstyle{achemso}
%\bibliography{referencesSiC}

\begin{mcitethebibliography}{55}
\providecommand*\natexlab[1]{#1}
\providecommand*\mciteSetBstSublistMode[1]{}
\providecommand*\mciteSetBstMaxWidthForm[2]{}
\providecommand*\mciteBstWouldAddEndPuncttrue
  {\def\EndOfBibitem{\unskip.}}
\providecommand*\mciteBstWouldAddEndPunctfalse
  {\let\EndOfBibitem\relax}
\providecommand*\mciteSetBstMidEndSepPunct[3]{}
\providecommand*\mciteSetBstSublistLabelBeginEnd[3]{}
\providecommand*\EndOfBibitem{}
\mciteSetBstSublistMode{f}
\mciteSetBstMaxWidthForm{subitem}{(\alph{mcitesubitemcount})}
\mciteSetBstSublistLabelBeginEnd
  {\mcitemaxwidthsubitemform\space}
  {\relax}
  {\relax}

\bibitem[Christle \latin{et~al.}(2015)Christle, Falk, Andrich, Klimov, Hassan,
  Son, Janzén, Ohshima, and Awschalom]{Christle2015}
Christle,~D.~J.; Falk,~A.~L.; Andrich,~P.; Klimov,~P.~V.; Hassan,~J.~U.;
  Son,~N.~T.; Janz\'{e}n,~E.; Ohshima,~T.; Awschalom,~D.~D. Isolated electron
  spins in silicon carbide with millisecond coherence times. \emph{Nature
  Materials} \textbf{2015}, \emph{14}, 160--163\relax
\mciteBstWouldAddEndPuncttrue
\mciteSetBstMidEndSepPunct{\mcitedefaultmidpunct}
{\mcitedefaultendpunct}{\mcitedefaultseppunct}\relax
\EndOfBibitem
\bibitem[Fuchs \latin{et~al.}(2015)Fuchs, Stender, Trupke, Simin, Pflaum,
  Dyakonov, and Astakhov]{Fuchs2015}
Fuchs,~F.; Stender,~B.; Trupke,~M.; Simin,~D.; Pflaum,~J.; Dyakonov,~V.;
  Astakhov,~G.~V. Engineering near-infrared single-photon emitters with
  optically active spins in ultrapure silicon carbide. \emph{Nature
  Communications} \textbf{2015}, \emph{6}, 7578\relax
\mciteBstWouldAddEndPuncttrue
\mciteSetBstMidEndSepPunct{\mcitedefaultmidpunct}
{\mcitedefaultendpunct}{\mcitedefaultseppunct}\relax
\EndOfBibitem
\bibitem[Falk \latin{et~al.}(2013)Falk, Buckley, Calusine, Koehl, Dobrovitski,
  Politi, Zorman, Feng, and Awschalom]{falk2013polytype}
Falk,~A.~L.; Buckley,~B.~B.; Calusine,~G.; Koehl,~W.~F.; Dobrovitski,~V.~V.;
  Politi,~A.; Zorman,~C.~A.; Feng,~P. X.-L.; Awschalom,~D.~D. Polytype control
  of spin qubits in silicon carbide. \emph{Nature Communications}
  \textbf{2013}, \emph{4}, 1--7\relax
\mciteBstWouldAddEndPuncttrue
\mciteSetBstMidEndSepPunct{\mcitedefaultmidpunct}
{\mcitedefaultendpunct}{\mcitedefaultseppunct}\relax
\EndOfBibitem
\bibitem[Widmann \latin{et~al.}(2015)Widmann, Lee, Rendler, Son, Fedder, Paik,
  Yang, Zhao, Yang, Booker, \latin{et~al.} others]{widmann2015coherentRT}
Widmann,~M.; Lee,~S.-Y.; Rendler,~T.; Son,~N.~T.; Fedder,~H.; Paik,~S.;
  Yang,~L.-P.; Zhao,~N.; Yang,~S.; Booker,~I.; others Coherent control of
  single spins in silicon carbide at room temperature. \emph{Nature Materials}
  \textbf{2015}, \emph{14}, 164--168\relax
\mciteBstWouldAddEndPuncttrue
\mciteSetBstMidEndSepPunct{\mcitedefaultmidpunct}
{\mcitedefaultendpunct}{\mcitedefaultseppunct}\relax
\EndOfBibitem
\bibitem[Seo \latin{et~al.}(2016)Seo, Falk, Klimov, Miao, Galli, and
  Awschalom]{Seo2016}
Seo,~H.; Falk,~A.~L.; Klimov,~P.~V.; Miao,~K.~C.; Galli,~G.; Awschalom,~D.~D.
  Quantum decoherence dynamics of divacancy spins in silicon carbide.
  \emph{Nature Communications} \textbf{2016}, \emph{7}, 12935\relax
\mciteBstWouldAddEndPuncttrue
\mciteSetBstMidEndSepPunct{\mcitedefaultmidpunct}
{\mcitedefaultendpunct}{\mcitedefaultseppunct}\relax
\EndOfBibitem
\bibitem[Nagy \latin{et~al.}(2019)Nagy, Niethammer, Widmann, Chen, Udvarhelyi,
  Bonato, Hassan, Karhu, Ivanov, Son, Maze, Ohshima, Soykal, Gali, Lee, Kaiser,
  and Wrachtrup]{nagy2019highfidelity}
Nagy,~R.; Niethammer,~M.; Widmann,~M.; Chen,~Y.-C.; Udvarhelyi,~P.; Bonato,~C.;
  Hassan,~J.~U.; Karhu,~R.; Ivanov,~I.~G.; Son,~N.~T.; Maze,~J.~R.;
  Ohshima,~T.; Soykal,~{\"O}.~O.; Gali,~{\'A}.; Lee,~S.-Y.; Kaiser,~F.;
  Wrachtrup,~J. High-fidelity spin and optical control of single
  silicon-vacancy centres in silicon carbide. \emph{Nature Communications}
  \textbf{2019}, \emph{10}, 1--8\relax
\mciteBstWouldAddEndPuncttrue
\mciteSetBstMidEndSepPunct{\mcitedefaultmidpunct}
{\mcitedefaultendpunct}{\mcitedefaultseppunct}\relax
\EndOfBibitem
\bibitem[Carter \latin{et~al.}(2015)Carter, Soykal, Dev, Economou, and
  Glaser]{Carter2015}
Carter,~S.~G.; Soykal,~O.~O.; Dev,~P.; Economou,~S.~E.; Glaser,~E.~R. Spin
  coherence and echo modulation of the silicon vacancy in
  $4H\ensuremath{-}\mathrm{SiC}$ at room temperature. \emph{Physical Review B}
  \textbf{2015}, \emph{92}, 161202\relax
\mciteBstWouldAddEndPuncttrue
\mciteSetBstMidEndSepPunct{\mcitedefaultmidpunct}
{\mcitedefaultendpunct}{\mcitedefaultseppunct}\relax
\EndOfBibitem
\bibitem[Gadalla \latin{et~al.}(2021)Gadalla, Greenspon, Defo, Zhang, and
  Hu]{gadalla2021enhanced}
Gadalla,~M.~N.; Greenspon,~A.~S.; Defo,~R.~K.; Zhang,~X.; Hu,~E.~L. Enhanced
  cavity coupling to silicon vacancies in 4H silicon carbide using laser
  irradiation and thermal annealing. \emph{Proceedings of the National Academy
  of Sciences} \textbf{2021}, \emph{118}\relax
\mciteBstWouldAddEndPuncttrue
\mciteSetBstMidEndSepPunct{\mcitedefaultmidpunct}
{\mcitedefaultendpunct}{\mcitedefaultseppunct}\relax
\EndOfBibitem
\bibitem[Majety \latin{et~al.}(2021)Majety, Norman, Li, Bell, Saha, and
  Radulaski]{majety2021quantum}
Majety,~S.; Norman,~V.~A.; Li,~L.; Bell,~M.; Saha,~P.; Radulaski,~M. Quantum
  photonics in triangular-cross-section nanodevices in silicon carbide.
  \emph{Journal of Physics: Photonics} \textbf{2021}, \emph{3}, 034008\relax
\mciteBstWouldAddEndPuncttrue
\mciteSetBstMidEndSepPunct{\mcitedefaultmidpunct}
{\mcitedefaultendpunct}{\mcitedefaultseppunct}\relax
\EndOfBibitem
\bibitem[Crook \latin{et~al.}(2020)Crook, Anderson, Miao, Bourassa, Lee,
  Bayliss, Bracher, Zhang, Abe, Ohshima, \latin{et~al.}
  others]{crookEHu2020purcell}
Crook,~A.~L.; Anderson,~C.~P.; Miao,~K.~C.; Bourassa,~A.; Lee,~H.;
  Bayliss,~S.~L.; Bracher,~D.~O.; Zhang,~X.; Abe,~H.; Ohshima,~T.; others
  Purcell enhancement of a single silicon carbide color center with coherent
  spin control. \emph{Nano Letters} \textbf{2020}, \emph{20}, 3427--3434\relax
\mciteBstWouldAddEndPuncttrue
\mciteSetBstMidEndSepPunct{\mcitedefaultmidpunct}
{\mcitedefaultendpunct}{\mcitedefaultseppunct}\relax
\EndOfBibitem
\bibitem[Bracher \latin{et~al.}(2017)Bracher, Zhang, and
  Hu]{bracherEHu2017selective}
Bracher,~D.~O.; Zhang,~X.; Hu,~E.~L. Selective Purcell enhancement of two
  closely linked zero-phonon transitions of a silicon carbide color center.
  \emph{Proceedings of the National Academy of Sciences} \textbf{2017},
  \emph{114}, 4060--4065\relax
\mciteBstWouldAddEndPuncttrue
\mciteSetBstMidEndSepPunct{\mcitedefaultmidpunct}
{\mcitedefaultendpunct}{\mcitedefaultseppunct}\relax
\EndOfBibitem
\bibitem[Nagy \latin{et~al.}(2018)Nagy, Widmann, Niethammer, Dasari, Gerhardt,
  Soykal, Radulaski, Ohshima, Vu{\v{c}}kovi{\'c}, Son, \latin{et~al.}
  others]{nagy2018quantumdichroic}
Nagy,~R.; Widmann,~M.; Niethammer,~M.; Dasari,~D.~B.; Gerhardt,~I.;
  Soykal,~{\"O}.~O.; Radulaski,~M.; Ohshima,~T.; Vu{\v{c}}kovi{\'c},~J.;
  Son,~N.~T.; others Quantum properties of dichroic silicon vacancies in
  silicon carbide. \emph{Physical Review Applied} \textbf{2018}, \emph{9},
  034022\relax
\mciteBstWouldAddEndPuncttrue
\mciteSetBstMidEndSepPunct{\mcitedefaultmidpunct}
{\mcitedefaultendpunct}{\mcitedefaultseppunct}\relax
\EndOfBibitem
\bibitem[Lukin \latin{et~al.}(2020)Lukin, Dory, Guidry, Yang, Mishra, Trivedi,
  Radulaski, Sun, Vercruysse, Ahn, \latin{et~al.} others]{lukinRadulaski20204h}
Lukin,~D.~M.; Dory,~C.; Guidry,~M.~A.; Yang,~K.~Y.; Mishra,~S.~D.; Trivedi,~R.;
  Radulaski,~M.; Sun,~S.; Vercruysse,~D.; Ahn,~G.~H.; others
  4H-silicon-carbide-on-insulator for integrated quantum and nonlinear
  photonics. \emph{Nature Photonics} \textbf{2020}, \emph{14}, 330--334\relax
\mciteBstWouldAddEndPuncttrue
\mciteSetBstMidEndSepPunct{\mcitedefaultmidpunct}
{\mcitedefaultendpunct}{\mcitedefaultseppunct}\relax
\EndOfBibitem
\bibitem[Radulaski \latin{et~al.}(2017)Radulaski, Widmann, Niethammer, Zhang,
  Lee, Rendler, Lagoudakis, Son, Janzen, Ohshima, \latin{et~al.}
  others]{radulaski2017scalable}
Radulaski,~M.; Widmann,~M.; Niethammer,~M.; Zhang,~J.~L.; Lee,~S.-Y.;
  Rendler,~T.; Lagoudakis,~K.~G.; Son,~N.~T.; Janzen,~E.; Ohshima,~T.; others
  Scalable quantum photonics with single color centers in silicon carbide.
  \emph{Nano Letters} \textbf{2017}, \emph{17}, 1782--1786\relax
\mciteBstWouldAddEndPuncttrue
\mciteSetBstMidEndSepPunct{\mcitedefaultmidpunct}
{\mcitedefaultendpunct}{\mcitedefaultseppunct}\relax
\EndOfBibitem
\bibitem[Yuan \latin{et~al.}(2020)Yuan, Fitzpatrick, Rodgers, Sangtawesin,
  Srinivasan, and de~Leon]{YuanDeLeon2020Charge}
Yuan,~Z.; Fitzpatrick,~M.; Rodgers,~L. V.~H.; Sangtawesin,~S.; Srinivasan,~S.;
  de~Leon,~N.~P. Charge state dynamics and optically detected electron spin
  resonance contrast of shallow nitrogen-vacancy centers in diamond.
  \emph{Physical Review Research} \textbf{2020}, \emph{2}, 033263\relax
\mciteBstWouldAddEndPuncttrue
\mciteSetBstMidEndSepPunct{\mcitedefaultmidpunct}
{\mcitedefaultendpunct}{\mcitedefaultseppunct}\relax
\EndOfBibitem
\bibitem[Sangtawesin \latin{et~al.}(2019)Sangtawesin, Dwyer, Srinivasan,
  Allred, Rodgers, De~Greve, Stacey, Dontschuk, O?Donnell, Hu, \latin{et~al.}
  others]{sangtawesinDeLeon2019origins}
Sangtawesin,~S.; Dwyer,~B.~L.; Srinivasan,~S.; Allred,~J.~J.; Rodgers,~L.~V.;
  De~Greve,~K.; Stacey,~A.; Dontschuk,~N.; O?Donnell,~K.~M.; Hu,~D.; others
  Origins of diamond surface noise probed by correlating single-spin
  measurements with surface spectroscopy. \emph{Physical Review X}
  \textbf{2019}, \emph{9}, 031052\relax
\mciteBstWouldAddEndPuncttrue
\mciteSetBstMidEndSepPunct{\mcitedefaultmidpunct}
{\mcitedefaultendpunct}{\mcitedefaultseppunct}\relax
\EndOfBibitem
\bibitem[V{\'a}squez \latin{et~al.}(2020)V{\'a}squez, Bathen, Galeckas,
  Bazioti, Johansen, Maestre, Cremades, Prytz, Moe, Kuznetsov, and
  Vines]{nanostr_VSi_expt_2020}
V{\'a}squez,~G.~C.; Bathen,~M.~E.; Galeckas,~A.; Bazioti,~C.; Johansen,~K.~M.;
  Maestre,~D.; Cremades,~A.; Prytz,~{\O}.; Moe,~A.~M.; Kuznetsov,~A.~Y.;
  Vines,~L. Strain Modulation of Si Vacancy Emission from SiC Micro- and
  Nanoparticles. \emph{Nano Letters} \textbf{2020}, \emph{20}, 8689--8695\relax
\mciteBstWouldAddEndPuncttrue
\mciteSetBstMidEndSepPunct{\mcitedefaultmidpunct}
{\mcitedefaultendpunct}{\mcitedefaultseppunct}\relax
\EndOfBibitem
\bibitem[Neethirajan \latin{et~al.}(2023)Neethirajan, Hache, Paone, Pinto,
  Denisenko, Stöhr, Udvarhelyi, Pershin, Gali, Wrachtrup, Kern, and
  Singha]{neethirajan2023}
Neethirajan,~J.~N.; Hache,~T.; Paone,~D.; Pinto,~D.; Denisenko,~A.; St\"{o}hr,~R.;
  Udvarhelyi,~P.; Pershin,~A.; Gali,~A.; Wrachtrup,~J.; Kern,~K.; Singha,~A.
  Controlled Surface Modification to Revive Shallow NV$^{-}$ Centers. \emph{Nano
  Letters} \textbf{2023}, \emph{23}, 2563--2569\relax
\mciteBstWouldAddEndPuncttrue
\mciteSetBstMidEndSepPunct{\mcitedefaultmidpunct}
{\mcitedefaultendpunct}{\mcitedefaultseppunct}\relax
\EndOfBibitem
\bibitem[Kaviani \latin{et~al.}(2014)Kaviani, Deák, Aradi, Frauenheim, Chou,
  and Gali]{Kaviani2014}
Kaviani,~M.; De\'{a}k,~P.; Aradi,~B.; Frauenheim,~T.; Chou,~J.-P.; Gali,~A. Proper
  Surface Termination for Luminescent Near-Surface NV Centers in Diamond.
  \emph{Nano Letters} \textbf{2014}, \emph{14}, 4772--4777\relax
\mciteBstWouldAddEndPuncttrue
\mciteSetBstMidEndSepPunct{\mcitedefaultmidpunct}
{\mcitedefaultendpunct}{\mcitedefaultseppunct}\relax
\EndOfBibitem
\bibitem[Li \latin{et~al.}(2019)Li, Chou, Wei, Sun, Hu, and Gali]{LiGali2019}
Li,~S.; Chou,~J.-P.; Wei,~J.; Sun,~M.; Hu,~A.; Gali,~A. Oxygenated (113)
  diamond surface for nitrogen-vacancy quantum sensors with preferential
  alignment and long coherence time from first principles. \emph{Carbon}
  \textbf{2019}, \emph{145}, 273--280\relax
\mciteBstWouldAddEndPuncttrue
\mciteSetBstMidEndSepPunct{\mcitedefaultmidpunct}
{\mcitedefaultendpunct}{\mcitedefaultseppunct}\relax
\EndOfBibitem
\bibitem[L\"{o}fgren \latin{et~al.}(2019)L\"{o}fgren, Pawar, Öberg, and
  Larsson]{Lofgren_2019}
L\"{o}fgren,~R.; Pawar,~R.; \"{O}berg,~S.; Larsson,~J.~A. The bulk conversion
  depth of the NV-center in diamond: computing a charged defect in a neutral
  slab. \emph{New Journal of Physics} \textbf{2019}, \emph{21}, 053037\relax
\mciteBstWouldAddEndPuncttrue
\mciteSetBstMidEndSepPunct{\mcitedefaultmidpunct}
{\mcitedefaultendpunct}{\mcitedefaultseppunct}\relax
\EndOfBibitem
\bibitem[Joshi and Dev(2022)Joshi, and Dev]{Joshi2022}
Joshi,~T.; Dev,~P. Site-Dependent Properties of Quantum Emitters in
  Nanostructured Silicon Carbide. \emph{PRX Quantum} \textbf{2022}, \emph{3},
  020325\relax
\mciteBstWouldAddEndPuncttrue
\mciteSetBstMidEndSepPunct{\mcitedefaultmidpunct}
{\mcitedefaultendpunct}{\mcitedefaultseppunct}\relax
\EndOfBibitem
\bibitem[Sayou~Ngomsi \latin{et~al.}(2024)Sayou~Ngomsi, Joshi, and
  Dev]{Sayou_PRMater2024}
Sayou~Ngomsi,~C.~A.; Joshi,~T.; Dev,~P. Optimum surface-passivation schemes for
  near-surface spin defects in silicon carbide. \emph{Phys. Rev. Mater.}
  \textbf{2024}, \emph{8}, 056202\relax
\mciteBstWouldAddEndPuncttrue
\mciteSetBstMidEndSepPunct{\mcitedefaultmidpunct}
{\mcitedefaultendpunct}{\mcitedefaultseppunct}\relax
\EndOfBibitem
\bibitem[Struck and D'Evelyn(1993)Struck, and D'Evelyn]{stability1}
Struck,~L.~M.; D'Evelyn,~M.~P. {Interaction of hydrogen and water with diamond
  (100): Infrared spectroscopy}. \emph{Journal of Vacuum Science \& Technology
  A} \textbf{1993}, \emph{11}, 1992--1997\relax
\mciteBstWouldAddEndPuncttrue
\mciteSetBstMidEndSepPunct{\mcitedefaultmidpunct}
{\mcitedefaultendpunct}{\mcitedefaultseppunct}\relax
\EndOfBibitem
\bibitem[Thomas \latin{et~al.}(1992)Thomas, Rudder, and Markunas]{stability2}
Thomas,~R.~E.; Rudder,~R.~A.; Markunas,~R.~J. {Thermal desorption from
  hydrogenated and oxygenated diamond (100) surfaces}. \emph{Journal of Vacuum
  Science \& Technology A} \textbf{1992}, \emph{10}, 2451--2457\relax
\mciteBstWouldAddEndPuncttrue
\mciteSetBstMidEndSepPunct{\mcitedefaultmidpunct}
{\mcitedefaultendpunct}{\mcitedefaultseppunct}\relax
\EndOfBibitem
\bibitem[Shinohara \latin{et~al.}(2005)Shinohara, Katagiri, Iwatsuji, Matsuda,
  Fujiyama, Kimura, and Niwano]{Shinohara2005}
Shinohara,~M.; Katagiri,~T.; Iwatsuji,~K.; Matsuda,~Y.; Fujiyama,~H.;
  Kimura,~Y.; Niwano,~M. Oxidation of the hydrogen terminated silicon surfaces
  by oxygen plasma investigated by in-situ infrared spectroscopy. \emph{Thin
  Solid Films} \textbf{2005}, \emph{475}, 128--132, Asian-European
  International Conference on Plasma Surface Engineering 2003 Proceedings of
  the 4th Asian-European International Conference on Plasma Surface
  Engineering\relax
\mciteBstWouldAddEndPuncttrue
\mciteSetBstMidEndSepPunct{\mcitedefaultmidpunct}
{\mcitedefaultendpunct}{\mcitedefaultseppunct}\relax
\EndOfBibitem
\bibitem[Polking \latin{et~al.}(2018)Polking, Dibos, de~Leon, and
  Park]{Polking_deLeon_SiC_surface2018}
Polking,~M.~J.; Dibos,~A.~M.; de~Leon,~N.~P.; Park,~H. Improving Defect-Based
  Quantum Emitters in Silicon Carbide via Inorganic Passivation. \emph{Advanced
  Materials} \textbf{2018}, \emph{30}, 1704543\relax
\mciteBstWouldAddEndPuncttrue
\mciteSetBstMidEndSepPunct{\mcitedefaultmidpunct}
{\mcitedefaultendpunct}{\mcitedefaultseppunct}\relax
\EndOfBibitem
\bibitem[Giannozzi \latin{et~al.}(2009)Giannozzi, Baroni, Bonini, Calandra,
  Car, Cavazzoni, Ceresoli, Chiarotti, Cococcioni, Dabo, {Dal Corso},
  de~Gironcoli, Fabris, Fratesi, Gebauer, Gerstmann, Gougoussis, Kokalj,
  Lazzeri, Martin-Samos, Marzari, Mauri, Mazzarello, Paolini, Pasquarello,
  Paulatto, Sbraccia, Scandolo, Sclauzero, Seitsonen, Smogunov, Umari, and
  Wentzcovitch]{QE-2009}
Giannozzi,~P.; Baroni,~S.; Bonini,~N.; Calandra,~M.; Car,~R.; Cavazzoni,~C.;
  Ceresoli,~D.; Chiarotti,~G.~L.; Cococcioni,~M.; Dabo,~I.; {Dal Corso},~A.;
  de~Gironcoli,~S.; Fabris,~S.; Fratesi,~G.; Gebauer,~R.; Gerstmann,~U.;
  Gougoussis,~C.; Kokalj,~A.; Lazzeri,~M.; Martin-Samos,~L.; Marzari,~N.;
  Mauri,~F.; Mazzarello,~R.; Paolini,~S.; Pasquarello,~A.; Paulatto,~L.;
  Sbraccia,~C.; Scandolo,~S.; Sclauzero,~G.; Seitsonen,~A.~P.; Smogunov,~A.;
  Umari,~P.; Wentzcovitch,~R.~M. QUANTUM ESPRESSO: a modular and open-source
  software project for quantum simulations of materials. \emph{Journal of
  Physics: Condensed Matter} \textbf{2009}, \emph{21}, 395502 (19pp)\relax
\mciteBstWouldAddEndPuncttrue
\mciteSetBstMidEndSepPunct{\mcitedefaultmidpunct}
{\mcitedefaultendpunct}{\mcitedefaultseppunct}\relax
\EndOfBibitem
\bibitem[Giannozzi \latin{et~al.}(2017)Giannozzi, Andreussi, Brumme, Bunau,
  Nardelli, Calandra, Car, Cavazzoni, Ceresoli, Cococcioni, Colonna, Carnimeo,
  Corso, de~Gironcoli, Delugas, Jr, Ferretti, Floris, Fratesi, Fugallo,
  Gebauer, Gerstmann, Giustino, Gorni, Jia, Kawamura, Ko, Kokalj,
  Küçükbenli, Lazzeri, Marsili, Marzari, Mauri, Nguyen, Nguyen, de-la Roza,
  Paulatto, Poncé, Rocca, Sabatini, Santra, Schlipf, Seitsonen, Smogunov,
  Timrov, Thonhauser, Umari, Vast, Wu, and Baroni]{QE-2017}
Giannozzi,~P.; Andreussi,~O.; Brumme,~T.; Bunau,~O.; Nardelli,~M.~B.;
  Calandra,~M.; Car,~R.; Cavazzoni,~C.; Ceresoli,~D.; Cococcioni,~M.;
  Colonna,~N.; Carnimeo,~I.; Corso,~A.~D.; de~Gironcoli,~S.; Delugas,~P.;
  Jr,~R. A.~D.; Ferretti,~A.; Floris,~A.; Fratesi,~G.; Fugallo,~G.;
  Gebauer,~R.; Gerstmann,~U.; Giustino,~F.; Gorni,~T.; Jia,~J.; Kawamura,~M.;
  Ko,~H.-Y.; Kokalj,~A.; K\"{u}\c{c}\"{u}kbenli,~E.; Lazzeri,~M.; Marsili,~M.;
  Marzari,~N.; Mauri,~F.; Nguyen,~N.~L.; Nguyen,~H.-V.; de-la Roza,~A.~O.;
  Paulatto,~L.; Ponc\'{e},~S.; Rocca,~D.; Sabatini,~R.; Santra,~B.; Schlipf,~M.;
  Seitsonen,~A.~P.; Smogunov,~A.; Timrov,~I.; Thonhauser,~T.; Umari,~P.;
  Vast,~N.; Wu,~X.; Baroni,~S. Advanced capabilities for materials modelling
  with QUANTUM ESPRESSO. \emph{Journal of Physics: Condensed Matter}
  \textbf{2017}, \emph{29}, 465901\relax
\mciteBstWouldAddEndPuncttrue
\mciteSetBstMidEndSepPunct{\mcitedefaultmidpunct}
{\mcitedefaultendpunct}{\mcitedefaultseppunct}\relax
\EndOfBibitem
\bibitem[Vanderbilt(1990)]{USPP_vanderbilt1990}
Vanderbilt,~D. Soft self-consistent pseudopotentials in a generalized
  eigenvalue formalism. \emph{Phys. Rev. B} \textbf{1990}, \emph{41},
  7892--7895\relax
\mciteBstWouldAddEndPuncttrue
\mciteSetBstMidEndSepPunct{\mcitedefaultmidpunct}
{\mcitedefaultendpunct}{\mcitedefaultseppunct}\relax
\EndOfBibitem
\bibitem[Perdew and Yue(1986)Perdew, and Yue]{GGA}
Perdew,~J.~P.; Yue,~W. Accurate and simple density functional for the
  electronic exchange energy: Generalized gradient approximation.
  \emph{Physical Review B} \textbf{1986}, \emph{33}, 8800--8802\relax
\mciteBstWouldAddEndPuncttrue
\mciteSetBstMidEndSepPunct{\mcitedefaultmidpunct}
{\mcitedefaultendpunct}{\mcitedefaultseppunct}\relax
\EndOfBibitem
\bibitem[Perdew \latin{et~al.}(1996)Perdew, Burke, and Ernzerhof]{PBE}
Perdew,~J.~P.; Burke,~K.; Ernzerhof,~M. Generalized Gradient Approximation Made
  Simple. \emph{Physical Review Letters} \textbf{1996}, \emph{77},
  3865--3868\relax
\mciteBstWouldAddEndPuncttrue
\mciteSetBstMidEndSepPunct{\mcitedefaultmidpunct}
{\mcitedefaultendpunct}{\mcitedefaultseppunct}\relax
\EndOfBibitem
\bibitem[Heyd \latin{et~al.}(2003)Heyd, Scuseria, and Ernzerhof]{HSE03}
Heyd,~J.; Scuseria,~G.~E.; Ernzerhof,~M. Hybrid functionals based on a screened
  Coulomb potential. \emph{The Journal of Chemical Physics} \textbf{2003},
  \emph{118}, 8207--8215\relax
\mciteBstWouldAddEndPuncttrue
\mciteSetBstMidEndSepPunct{\mcitedefaultmidpunct}
{\mcitedefaultendpunct}{\mcitedefaultseppunct}\relax
\EndOfBibitem
\bibitem[Heyd \latin{et~al.}(2006)Heyd, Scuseria, and Ernzerhof]{HSE06}
Heyd,~J.; Scuseria,~G.~E.; Ernzerhof,~M. {Erratum: Hybrid functionals based on
  a screened Coulomb potential [J. Chem. Phys. 118, 8207 (2003)]}. \emph{The
  Journal of Chemical Physics} \textbf{2006}, \emph{124}, 219906\relax
\mciteBstWouldAddEndPuncttrue
\mciteSetBstMidEndSepPunct{\mcitedefaultmidpunct}
{\mcitedefaultendpunct}{\mcitedefaultseppunct}\relax
\EndOfBibitem
\bibitem[Wolfowicz \latin{et~al.}(2020)Wolfowicz, Anderson, Diler, Poluektov,
  Heremans, and Awschalom]{Wolfowicz2020}
Wolfowicz,~G.; Anderson,~C.~P.; Diler,~B.; Poluektov,~O.~G.; Heremans,~F.~J.;
  Awschalom,~D.~D. Vanadium spin qubits as telecom quantum emitters in silicon
  carbide. \emph{Science Advances} \textbf{2020}, \emph{6}, eaaz1192\relax
\mciteBstWouldAddEndPuncttrue
\mciteSetBstMidEndSepPunct{\mcitedefaultmidpunct}
{\mcitedefaultendpunct}{\mcitedefaultseppunct}\relax
\EndOfBibitem
\bibitem[Breev \latin{et~al.}(2022)Breev, Shang, Poshakinskiy, Singh,
  Berenc{\'e}n, Hollenbach, Nagalyuk, Mokhov, Babunts, Baranov, Suter,
  Tarasenko, Astakhov, and Anisimov]{Breev2022}
Breev,~I.~D.; Shang,~Z.; Poshakinskiy,~A.~V.; Singh,~H.; Berenc{\'e}n,~Y.;
  Hollenbach,~M.; Nagalyuk,~S.~S.; Mokhov,~E.~N.; Babunts,~R.~A.;
  Baranov,~P.~G.; Suter,~D.; Tarasenko,~S.~A.; Astakhov,~G.~V.; Anisimov,~A.~N.
  Inverted fine structure of a 6H-SiC qubit enabling robust spin-photon
  interface. \emph{npj Quantum Information} \textbf{2022}, \emph{8}, 23\relax
\mciteBstWouldAddEndPuncttrue
\mciteSetBstMidEndSepPunct{\mcitedefaultmidpunct}
{\mcitedefaultendpunct}{\mcitedefaultseppunct}\relax
\EndOfBibitem
\bibitem[Carter \latin{et~al.}(2025)Carter, Tathfif, Burgess, VanMil, Debata,
  and Dev]{CarterDev2025}
Carter,~S.~G.; Tathfif,~I.; Burgess,~C.; VanMil,~B.; Debata,~S.; Dev,~P.
  Influence of nitrogen doping and annealing on the silicon vacancy in
  $4H\text{\ensuremath{-}}\mathrm{SiC}$. \emph{Phys. Rev. B} \textbf{2025},
  \emph{112}, 085209\relax
\mciteBstWouldAddEndPuncttrue
\mciteSetBstMidEndSepPunct{\mcitedefaultmidpunct}
{\mcitedefaultendpunct}{\mcitedefaultseppunct}\relax
\EndOfBibitem
\bibitem[Stuermer \latin{et~al.}(2025)Stuermer, Steidl, Woernle, Schustereder,
  Zmoelnig, Riss, Schoenherr, Urlesberger, Edelmann, Kaschel, Kern, Anders,
  Krainer, Heiss, and Wrachtrup]{Stuermer2025}
Stuermer,~P.~A.; Steidl,~T.; Woernle,~R.; Schustereder,~W.; Zmoelnig,~C.;
  Riss,~J.; Schoenherr,~H.; Urlesberger,~P.; Edelmann,~K.; Kaschel,~M.;
  Kern,~M.; Anders,~J.; Krainer,~S.; Heiss,~H.; Wrachtrup,~J. Low loss 4H-SiC
  photonics for spin-quantum sensing. \emph{Applied Physics Letters}
  \textbf{2025}, \emph{127}, 104001\relax
\mciteBstWouldAddEndPuncttrue
\mciteSetBstMidEndSepPunct{\mcitedefaultmidpunct}
{\mcitedefaultendpunct}{\mcitedefaultseppunct}\relax
\EndOfBibitem
\bibitem[Monkhorst and Pack(1976)Monkhorst, and Pack]{Monkhorst}
Monkhorst,~H.~J.; Pack,~J.~D. Special points for Brillouin-zone integrations.
  \emph{Physical Review B} \textbf{1976}, \emph{13}, 5188--5192\relax
\mciteBstWouldAddEndPuncttrue
\mciteSetBstMidEndSepPunct{\mcitedefaultmidpunct}
{\mcitedefaultendpunct}{\mcitedefaultseppunct}\relax
\EndOfBibitem
\bibitem[Gali \latin{et~al.}(2009)Gali, Janz\'en, De\'ak, Kresse, and
  Kaxiras]{KaxirasNV}
Gali,~A.; Janz\'en,~E.; De\'ak,~P.; Kresse,~G.; Kaxiras,~E. Theory of
  Spin-Conserving Excitation of the $N\ensuremath{-}{V}^{\ensuremath{-}}$
  Center in Diamond. \emph{Phys. Rev. Lett.} \textbf{2009}, \emph{103},
  186404\relax
\mciteBstWouldAddEndPuncttrue
\mciteSetBstMidEndSepPunct{\mcitedefaultmidpunct}
{\mcitedefaultendpunct}{\mcitedefaultseppunct}\relax
\EndOfBibitem
\bibitem[Dev(2020)]{PDEV_hBN_2020}
Dev,~P. Fingerprinting quantum emitters in hexagonal boron nitride using
  strain. \emph{Physical Review Research} \textbf{2020}, \emph{2},
  022050(R)\relax
\mciteBstWouldAddEndPuncttrue
\mciteSetBstMidEndSepPunct{\mcitedefaultmidpunct}
{\mcitedefaultendpunct}{\mcitedefaultseppunct}\relax
\EndOfBibitem
\bibitem[Narayanan and Dev(2023)Narayanan, and Dev]{narayanan2023}
Narayanan,~S.~K.; Dev,~P. Substrate-Induced Modulation of Quantum Emitter
  Properties in 2D Hexagonal Boron Nitride: Implications for Defect-Based
  Single Photon Sources in 2D Layers. \emph{ACS Applied Nano Materials}
  \textbf{2023}, \emph{6}, 3446--3452\relax
\mciteBstWouldAddEndPuncttrue
\mciteSetBstMidEndSepPunct{\mcitedefaultmidpunct}
{\mcitedefaultendpunct}{\mcitedefaultseppunct}\relax
\EndOfBibitem
\bibitem[Huang \latin{et~al.}(2008)Huang, Sun, Tao, Menard, Nuzzo, and
  Zuo]{Huang2008SkinEffect}
Huang,~W.~J.; Sun,~R.; Tao,~J.; Menard,~L.~D.; Nuzzo,~R.~G.; Zuo,~J.~M.
  Coordination-dependent surface atomic contraction in nanocrystals revealed by
  coherent diffraction. \emph{Nature Materials} \textbf{2008}, \emph{7},
  308--313\relax
\mciteBstWouldAddEndPuncttrue
\mciteSetBstMidEndSepPunct{\mcitedefaultmidpunct}
{\mcitedefaultendpunct}{\mcitedefaultseppunct}\relax
\EndOfBibitem
\bibitem[Shiraishi(1990)]{Shiraishi_1990}
Shiraishi,~K. A New Slab Model Approach for Electronic Structure Calculation of
  Polar Semiconductor Surface. \emph{Journal of the Physical Society of Japan}
  \textbf{1990}, \emph{59}, 3455--3458\relax
\mciteBstWouldAddEndPuncttrue
\mciteSetBstMidEndSepPunct{\mcitedefaultmidpunct}
{\mcitedefaultendpunct}{\mcitedefaultseppunct}\relax
\EndOfBibitem
\bibitem[Deng \latin{et~al.}(2012)Deng, Li, Li, and Wei]{SuHuaiWei_PH_PRB}
Deng,~H.-X.; Li,~S.-S.; Li,~J.; Wei,~S.-H. Effect of hydrogen passivation on
  the electronic structure of ionic semiconductor nanostructures. \emph{Phys.
  Rev. B} \textbf{2012}, \emph{85}, 195328\relax
\mciteBstWouldAddEndPuncttrue
\mciteSetBstMidEndSepPunct{\mcitedefaultmidpunct}
{\mcitedefaultendpunct}{\mcitedefaultseppunct}\relax
\EndOfBibitem
\bibitem[Yoo \latin{et~al.}(2021)Yoo, Todorova, Wickramaratne, Weston, Walle,
  and Neugebauer]{Yoo2021}
Yoo,~S.-H.; Todorova,~M.; Wickramaratne,~D.; Weston,~L.; Walle,~C. G. V.~d.;
  Neugebauer,~J. Finite-size correction for slab supercell calculations of
  materials with spontaneous polarization. \emph{npj Computational Materials}
  \textbf{2021}, \emph{7}, 58\relax
\mciteBstWouldAddEndPuncttrue
\mciteSetBstMidEndSepPunct{\mcitedefaultmidpunct}
{\mcitedefaultendpunct}{\mcitedefaultseppunct}\relax
\EndOfBibitem
\bibitem[Choi \latin{et~al.}(2005)Choi, Puthenkovilakam, and
  Chang]{Choi_band_alignment2005}
Choi,~J.; Puthenkovilakam,~R.; Chang,~J.~P. Band structure and alignment of the
  AlN/SiC heterostructure. \emph{Applied Physics Letters} \textbf{2005},
  \emph{86}, 192101\relax
\mciteBstWouldAddEndPuncttrue
\mciteSetBstMidEndSepPunct{\mcitedefaultmidpunct}
{\mcitedefaultendpunct}{\mcitedefaultseppunct}\relax
\EndOfBibitem
\bibitem[Freysoldt \latin{et~al.}(2009)Freysoldt, Neugebauer, and Van~de
  Walle]{FNV_2009}
Freysoldt,~C.; Neugebauer,~J.; Van~de Walle,~C.~G. Fully Ab Initio Finite-Size
  Corrections for Charged-Defect Supercell Calculations. \emph{Phys. Rev.
  Lett.} \textbf{2009}, \emph{102}, 016402\relax
\mciteBstWouldAddEndPuncttrue
\mciteSetBstMidEndSepPunct{\mcitedefaultmidpunct}
{\mcitedefaultendpunct}{\mcitedefaultseppunct}\relax
\EndOfBibitem
\bibitem[Naik and Jain(2018)Naik, and Jain]{coffee_2018}
Naik,~M.~H.; Jain,~M. CoFFEE: Corrections For Formation Energy and Eigenvalues
  for charged defect simulations. \emph{Computer Physics Communications}
  \textbf{2018}, \emph{226}, 114--126\relax
\mciteBstWouldAddEndPuncttrue
\mciteSetBstMidEndSepPunct{\mcitedefaultmidpunct}
{\mcitedefaultendpunct}{\mcitedefaultseppunct}\relax
\EndOfBibitem
\bibitem[Soykal \latin{et~al.}(2016)Soykal, Dev, and
  Economou]{soykal2016silicon}
Soykal,~{\"O}.; Dev,~P.; Economou,~S.~E. Silicon vacancy center in 4 H-SiC:
  Electronic structure and spin-photon interfaces. \emph{Physical Review B}
  \textbf{2016}, \emph{93}, 081207\relax
\mciteBstWouldAddEndPuncttrue
\mciteSetBstMidEndSepPunct{\mcitedefaultmidpunct}
{\mcitedefaultendpunct}{\mcitedefaultseppunct}\relax
\EndOfBibitem
\bibitem[Economou and Dev(2016)Economou, and Dev]{economou2016spin}
Economou,~S.~E.; Dev,~P. Spin-photon entanglement interfaces in silicon carbide
  defect centers. \emph{Nanotechnology} \textbf{2016}, \emph{27}, 504001\relax
\mciteBstWouldAddEndPuncttrue
\mciteSetBstMidEndSepPunct{\mcitedefaultmidpunct}
{\mcitedefaultendpunct}{\mcitedefaultseppunct}\relax
\EndOfBibitem
\bibitem[Dev \latin{et~al.}(2008)Dev, Xue, and Zhang]{Dev_PRL_DeepDefects_2008}
Dev,~P.; Xue,~Y.; Zhang,~P. {Defect-Induced Intrinsic Magnetism in Wide-Gap III
  Nitrides}. \emph{Physical Review Letters} \textbf{2008}, \emph{100},
  117204\relax
\mciteBstWouldAddEndPuncttrue
\mciteSetBstMidEndSepPunct{\mcitedefaultmidpunct}
{\mcitedefaultendpunct}{\mcitedefaultseppunct}\relax
\EndOfBibitem
\bibitem[Dev and Zhang(2010)Dev, and Zhang]{Dev_PRB_DeepDefects_2010}
Dev,~P.; Zhang,~P. Unconventional magnetism in semiconductors: Role of
  localized acceptor states. \emph{Physical Review B} \textbf{2010}, \emph{81},
  085207\relax
\mciteBstWouldAddEndPuncttrue
\mciteSetBstMidEndSepPunct{\mcitedefaultmidpunct}
{\mcitedefaultendpunct}{\mcitedefaultseppunct}\relax
\EndOfBibitem
\bibitem[Dev \latin{et~al.}(2010)Dev, Zeng, and Zhang]{Dev_PRB_NW_2010}
Dev,~P.; Zeng,~H.; Zhang,~P. Defect-induced magnetism in nitride and oxide
  nanowires: Surface effects and quantum confinement. \emph{Physical Review B}
  \textbf{2010}, \emph{82}, 165319\relax
\mciteBstWouldAddEndPuncttrue
\mciteSetBstMidEndSepPunct{\mcitedefaultmidpunct}
{\mcitedefaultendpunct}{\mcitedefaultseppunct}\relax
\EndOfBibitem
\end{mcitethebibliography}
\providecommand{\latin}[1]{#1}
\makeatletter
\providecommand{\doi}
  {\begingroup\let\do\@makeother\dospecials
  \catcode`\{=1 \catcode`\}=2 \doi@aux}
\providecommand{\doi@aux}[1]{\endgroup\texttt{#1}}
\makeatother
\providecommand*\mcitethebibliography{\thebibliography}
\csname @ifundefined\endcsname{endmcitethebibliography}
  {\let\endmcitethebibliography\endthebibliography}{}

\clearpage

\begin{figure}[!]
\centering
\includegraphics[width=8.25cm]{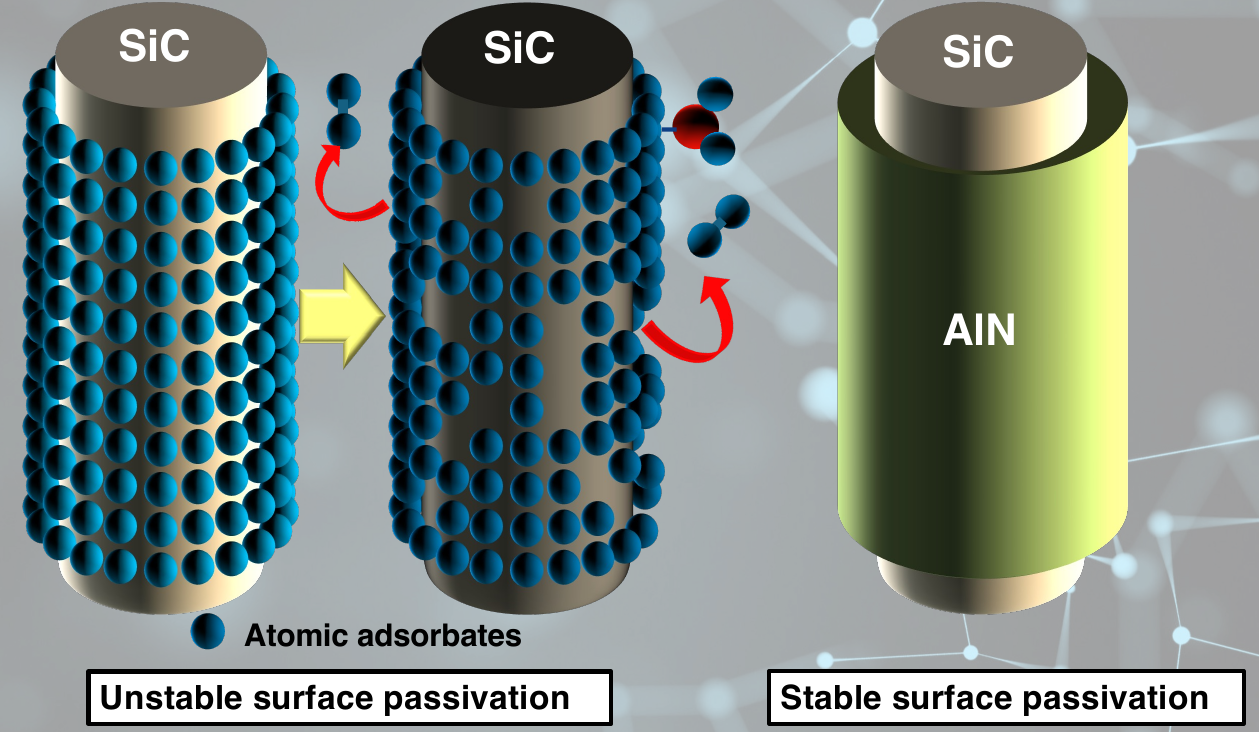}\par
Table of Contents Graphic
\end{figure}

\end{document}